\documentclass[sigconf]{acmart}
\AtBeginDocument{%
  }

\usepackage{multirow} 
\usepackage{tabularx} 
\usepackage{graphicx}
\usepackage{xcolor}
\usepackage[table]{xcolor}
\usepackage{makecell} 
\usepackage{float}  
\usepackage{placeins} 
\usepackage[utf8]{inputenc}
\usepackage[most]{tcolorbox}
\tcbuselibrary{listingsutf8}
\usepackage{setspace}
\usepackage{array}

\usepackage{amssymb}

\renewcommand\footnotetextcopyrightpermission[1]{} 
\setcopyright{none} 

\begin{document}

\title{Social Knowledge for Cross-Domain User Preference Modeling}

\newcommand{\tsc}[1]{\textsuperscript{#1}} 
\author{Nir Lotan\tsc{1}, Adir Solomon\tsc{1}, Ido Guy\tsc{2}, Einat Minkov\tsc{1}} 
\affiliation{%
  \institution{1. University of Haifa}
  \institution{2. Ben-Gurion University of the Negev}
    \country{Israel}
}

\renewcommand{\shortauthors}{Lotan et al.}

\begin{abstract}
In this work, we demonstrate that user preferences can be represented and predicted across topical domains using large-scale social modeling. Given information about popular entities favored by a user, we project the user into a social embedding space learned from a large-scale sample of the Twitter (now X) network. By representing both users and popular entities in a joint social space, we can assess the relevance of candidate entities (e.g., music artists) using cosine similarity within this embedding space. A comprehensive evaluation using link prediction experiments shows that this method achieves effective personalization even when no user feedback is available for items in the target domain, yielding substantial improvements over a strong popularity-based baseline. Further in-depth analysis illustrates that socio-demographic factors encoded in the social embeddings are correlated with user preferences across domains. Finally, we argue and demonstrate that the proposed approach can facilitate social modeling of end users using large language models (LLMs).
\end{abstract}



\maketitle
\pagestyle{plain}

\section{Introduction}

Personalization is a desired feature of AI systems when interacting with end users. In particular, recommender systems are inherently personalized, as their goal is to rank a set of candidate items, such as movies, according to an individual user’s preferences. To model these preferences, traditional recommender systems rely on explicit or implicit feedback indicating which items a user likes or dislikes. The relevance of new items is then estimated based on item similarity or through collaborative inference using large-scale user–item interaction data. However, this approach of user modeling faces several challenges. One major limitation is the need for a sufficient amount of feedback to accurately capture a user’s tastes. This implies that in `cold-start' scenarios, when there is only little feedback available, personalization is severely limited. Another drawback is that user feedback is typically domain-specific, e.g., limited to {\it movies}, and do not generalize to other domains, e.g., {\it car models}. In practice, however, user preferences are correlated with social factors, such as their {\it age}, {\it gender}, or {\it education level}. This implies that one's preferences across different domains are generally correlated. For example, a user who likes action movies may be more likely to favor sports cars. 

In this work, we turn to the social network of Twitter as a source of collective knowledge about user preferences across a wide range of topical domains. Twitter users typically associate themselves, aka {\it follow}, multiple accounts of popular entities in diverse domains, such as {\it musical artists}, {\it sports teams}, {\it news sources}, {\it politicians}, and more. Similar to a matrix of user-item ratings, such network information encodes latent patterns of user preferences. We utilize pre-trained low-dimension embeddings of popular Twitter accounts--referred to as {\it entities}--learned from a large sample of Twitter users and the accounts that they follow~\cite{lotanPLOS23}. 
Similar to word embeddings that capture word meaning based on word co-occurrences patterns in text~\cite{Mikolov2013}, the learned entity embeddings represent social semantics, having entities that individual Twitter users tend to co-follow reside close to each other in the social embedding space. It has been shown, for example, that the learned embeddings of news outlets capture political bias information, due to the tendency of users to co-follow multiple news sources that align with their own political camp~\cite{lotanPLOS23}. Likewise, the entities of luxury sports cars would be similar if the same users co-follow these entity accounts. 

In general, information about the accounts that a user links or interacts with is commonly used to model community structures and homophily among users. However, user embeddings are typically learned from small social graphs in a transductive fashion~\cite{islam2021analysis}, preventing generalization across new users or domains. In contrast, our approach lends itself to inductive social user modeling. Given information about some entities that some individual user favors, that user is projected into the social embedding space based on the pre-trained embeddings of those entities. This social network information is indicative of the users' beliefs, personal preferences and traits~\cite{culotta2015predicting,xiaoKDD20,muellerCSCW21,twhinKDD22}. Indeed, it has been shown that socio-demographic traits of users may be inferred from their social embeddings by means of supervised classification, including their {\it gender}, {\it age}, {\it ethnicity}, {\it education level}, and {\it political affiliation}~\cite{lotanPLOS23}. 

In this work, we exploit and assess this social user representation scheme towards cross-domain personalization. Concretely, given a user representation, comprised of the pre-trained embeddings of popular entities that she follows, we assess the similarity of other entity accounts of interest in terms of cosine similarity in the social embedding space. Our experiments assess cross-domain user preference prediction using a dedicated dataset which we constructed for this purpose. The dataset consists of the social representation of roughly 12K Twitter users, who are known to follow top-popular entities in 14 different topical domains. We assess personalization though the proxy task of link prediction. Concretely, we generate personalized rankings of candidate entities in each target domain in turn using social similarity to the users, where we consider those entities that each user is known to follow as relevant responses.  

Our experiments show that this method of social user representation achieves effective personalization, improving MAP performance by up to 22\% over the strong one-fit-all baseline of popularity-based ranking. These results hold also when no concrete user information in the target domain is available in their profile. We further show performance converges relatively fast as more entities are associated with each user, where as few as 10 entities per user already support cross-domain personalization. In addition to these qualitative experiments, we include qualitative analysis results, which illustrate the correlations that exist between socio-demographic factors that are encoded in the social user embeddings and their preferences, as observed in our experimental dataset. Finally, we examine the implications of this study for personalization in LLMs. Providing a list of entity names that the user likes, we follow the same evaluation protocol, prompting the model of GPT4o to generate personal rankings of the candidate entities. The LLM's results replicate our own findings, showing that specifying as few as 12 entities per user improves MAP  performance by 13\% compared to a one-fit-all ranking in restrictive settings, and performance improves when more example entities are provided. Thus, a main conclusion of this research is that users' preferences can be elicited implicitly from a small number of topical categories and entities of interest. As we discuss, such information may be easily obtained from new users using a lightweight entry point. 

In summary, this paper makes the following main contributions:
\begin{itemize}
\item We show that information about popular entities is an effective form of social user modeling, enabling the predictions of user preferences across topical domains.
\item We evaluate the inference of cross-domain personal preferences using a dataset constructed from real-world data of social media users. This dataset may be shared for research purposes.
\item The proposed social user modeling approach is shown to enable effective personalization in LLMs, requiring minimal inputs from new users.
\item We highlight the correlation between user preferences and a variety of socio-demographic traits. While beneficial, this also indicates that there are social biases that may be reinforced by social user modeling.
\end{itemize}

\section{Related work}

\subsection{User embeddings for personalization}

It is a common approach in recommender systems to factorize an observed user-item rating matrix into products of two sets of latent factors for users and items respectively. This collaborative filtering (CF) approach has shown great power in capturing user interests for achieving personalized recommendation, further leveraging dot-product of two factors to
predict potential ratings (Hu et al., 2008; Koren et al., 2009; Rendle et al., 2009; Srebro et al., 2004; Zheng et al., 2016). However, CF models learn user embedding
factors in a transductive fashion, making it difficult to handle new users on-the-fly. Various recent works therefore explore inductive collaborative filtering framework that can generalize to new users. For example, it has been proposed to identify `key users' to obtain meta user embeddings, and leverage these meta latents to inductively compute embeddings
for few-shot users with limited training ratings and new unseen users via neural message passing~\cite{wuICML21}. 

Another limitation of user embeddings learned by CF systems is that user-item interaction history is typically limited to a single domain. As discussed in recent surveys~\cite{xuWWW24,zhangTIST}, it is therefore desired to transfer pre-trained latent user and item features from the source domain towards another domain. It may be possible to establish connections between multiple domains through overlapping users; for example, some users may interact with both music and books. However, in real-world applications, the number of overlapping users across domains is typically a minority. Otherwise, similar item text descriptions or images may be utilized towards knowledge transfer. It is currently open questions what knowledge should be mined in each domain, and how to establish linkages between domains and realize the transfer of relevant knowledge.

\subsection{Social user embeddings}

In this work, we utilize pre-trained entity embeddings that were learned from large-scale social network data of Twitter to construct social user embeddings in inductive manner. A highly related work presented TwHIN--embedding of Twitter users and other entities for personalized recommendation. This work applied the TransE algorithm~\cite{transe} to learn network-based embeddings of users, tweets, advertisers and ads from a heterogeneous graph that represented these entities and the links between them at production scale~\cite{twhinKDD22}. They proposed to represent `out-of-vocabulary' entities, such as new users, as a mixture over the embeddings of existing users who shared similar network patterns. The evaluation of TwHIN focused on personalized recommendation and search in Twitter, for example, recommending ads to users based on their inter-similarity in the shared embedding space. The source data and the embeddings learned in their work are not publicly accessible, however. In our work, we utilize the public pre-trained embeddings of SocialVec~\cite{lotanPLOS23}, which pertain to roughly 200K popular accounts in Twitter, learned from a large sample of the Twitter network. Similar to their work, we represent individual users in inductive fashion, projecting the users onto the learned social embedding space based on the pre-trained embeddings of popular accounts which they follow. Unlike previous works, we show that this representation supports personalization, allowing to predict user preferences across multiple topical domains. We further report extensive experiments, showing that little evidence is needed to achieve effective personalization.

\subsection{Personalization in LLMs}

A related line of research employs LLMs as assistants for recommendation, having user preferences communicated in natural language. For example, a prompt may include example item descriptions along with user feedback--specifying at least 20 ratings per user~\cite{lyuNAACL24}, as well candidate item descriptions, having the model instructed to rate the candidate items by the user's preferences. Open challenges are the prompt size, which limits the number of historical feedbacks~\cite{liuACL24}. In addition, the textual description of items may be under-specified or vague~\cite{liuACL24,liangColing25}. While cross-domain recommendation might utilize similarity between textual descriptions of items, detailed user ratings are typically required for model finetuning~\citet{caoNAACL24}, where generalization to new tasks is non-trivial~\cite{wangNAACL24}. 

More generally, conversation agents typically track the dialogue history to generate customized responses tailored for each user. A lack of historical interactions therefore pose a challenge for new users. To that end, some techniques explicitly generate user preferences and profiles in natural language to augment LLMs’ input~\cite{Richardson2023}. Incorporating detailed user descriptions into the prompt for personalization purposes may lead to exceeding the context length however, as well as increased inference costs. some studies therefore apply retrieval methods to select the most relevant part of user behavior history to enhance LLM personalization~\cite{mysore24,salemiACL24}. A recent work which aimed to learn personalized model parameters that encapsulate user preferences, indicated the lack of comprehensive user modeling for this purpose. In particular, they claim that multiple user representations should be combined, as structured persona attributes, such as age and gender, are explicit but sparse, textual persona descriptions are noisy, and dialogue history query can be both noisy and uninformative for persona modeling~\cite{tanEMNLP24}.

Evaluation-wise, the evaluation of LLMs on personalized response generation remains relatively understudied~\cite{salemiACL24}. Various recent works proposed relevant benchmarks, which allow to simulate a large number of users, and assess the personalization of LLM models for applications such as scholarly search, text generation and recommendation tasks~\cite{salemiACL24,personallmICLR25}.

Compared with this existing research, we propose a novel paradigm for user representation for LLMs, describing the user as a list of popular entities that they like. We provide quantitative evaluation using a dataset of users extracted from Twitter. We believe that our study of personal user modeling using dedicated social embedding, learned outside of the `black box' of LLM, justifies and provides valuable insights about the information that is encoded in this form of user modeling. Furthermore, our dataset serves as a benchmark for user preference evaluation.

\section{Social user modeling}

Recommender systems typically model user preferences with respect to items to be purchased or consumed. We rather consider user preferences with respect to social {\it entities}, which maintain popular accounts on social media. Such entities may denote {\it musical artists}, {\it movies}, {\it TV shows}, {\it authors}, {\it news outlets}, and more. We make the following conjectures: (i) Personal preferences of entities are correlated with social factors, such as socio-demographic traits. Therefore, describing users based on  entities that they follow on social media would provide relevant evidence for
predicting their preferences towards other entities. (ii) In particular, we expect this form of social user modeling to generalize across topical domains. Once these conjectures are confirmed,  we address the question: How can users outside of social media be modeled using these principles? Namely, what is personalization capacity given a limited number of entities, for which users could provide feedback via a questionnaire?

In this section, we formally outline the social user modeling approach. Thereafter, Sections~\ref{sec:experiments}-~\ref{sec:llm} describe our experiments and analyses, addressing the stated conjectures and research questions.  

\subsection{Method}

Social media platforms contain numerous traces that reveal users’ personal traits and interests. Beyond content-based evidence--such as the text users post--personal characteristics are also reflected in network-based signals. In particular, prior research has shown that a wide range of personal traits can be inferred with high accuracy from the popular accounts users choose to follow on social media~\cite{culotta2015predicting,muellerCSCW21,lotanPLOS23}. For example, sports-related accounts are disproportionately followed by male users, while following the {\it New York Times} is a distinctive marker of highly educated users~\cite{lotanPLOS23}. More broadly, the collection of accounts a user follows provides insight into their interests, community affiliations, and geographical orientation~\cite{culotta2015predicting}.

It has been argued that similar to words, popular accounts on social media--henceforth, {\it entities}--may be treated as general vocabulary of social meaning. While word embeddings capture word meaning based on word co-occurrences in text, entity embeddings learned from social network data represent social semantics~\cite{twhinKDD22,lotanPLOS23}. In this work, we utilize publicly available pre-trained embeddings of  Twitter entities that were learned in the following fashion~\cite{lotanPLOS23}. 

Given on a large random sample of 1.5K Twitter users and the accounts that they follow, the most popular accounts--those with the highest number of followers within the sample--comprise a ‘vocabulary’ of 200K {\it entities} of general interest. The social embeddings of these entities were learned from the sampled network data, using an adaptation of the Word2Vec method~\cite{Mikolov2013}. Concretely, the well-known skip-gram variant of Word2Vec learns word embeddings by predicting neighboring words around a given focus word, assuming that local word sequences exhibit topical and syntactic coherence. In learning social entity embeddings, it is analogously assumed that the entities co-followed by individual users form coherent social contexts, as they all reflect the preferences and interests of those users. Accordingly, for each entity account followed by a user in the sampled data, the embedding parameters were tuned with the goal of predicting any of the other entities followed by the same user. Consequently, entities that are frequently co-followed by the same users are positioned close to one another in the social embedding space. The entity embeddings have been shown to capture both social and topical semantics. For instance, the representations of conservative news sources and politicians reside close to each in the embedding space, reflecting the tendency of conservative users to co-follow multiple entities that align with their views, and vice versa~\cite{lotanPLOS23}.

\paragraph*{Social user embeddings.} Given information about the entity accounts that an individual user $u_i$ follows, let the user be associated with the pre-trained embeddings of those entities, $u_i \rightarrow \{\vec{e}_{i}\}$. To project the user onto the embedding space, we follow the common practice of averaging the values of the bag-of-entity-embeddings $\{\vec{e}_{i}\}$ into a unified summary vector of the same dimension~\cite{shenACL18}. It has been shown that various socio-demographic traits, including {\it age}, {\it gender}, {\it ethnicity}, {\it education} level, and {\it political affiliation} can be inferred from the resulting user embeddings at high precision, using a classifier trained on labeled examples~\cite{lotanPLOS23}. We utilize those classifiers in our analyses below. 

An important characteristic of this user representation scheme is that it is inductive. Prior work has learned user embeddings from small-scale social graphs, primarily to model homophily among users and thereby promote generalization (e.g.,~\cite{islam2021analysis,sutterEACL24}). A key limitation of such transductive approaches is that they produce user representations that are specialized to a particular topic and restricted to a specific user population~\cite{twhinKDD22}. While alternative user embedding methods are plausible, we argue that any suitable approach must satisfy two essential requirements: (a) the embedding space should capture general dimensions of social meaning inferred from large-scale data; and (b) individual user representations should be constructed inductively from existing social embeddings, enabling transfer learning and generalization across users and domains.

\paragraph{Personal preferences prediction.}

In this work, we argue and demonstrate that once a user is projected into the social embedding space, the relevance of any entity to that user can be directly estimated via vector similarity~\cite{twhinKDD22,lotanPLOS23}. This enables personalized similarity assessments for entities and domains with which the user has not previously interacted, based solely on their social proximity to the user representation. Although Twitter does not provide structured semantic information about accounts—for example, whether an account represents a musical artist or a politician—we associate entities with topical categories for the purposes of this study. To obtain category information, we rely on the alignment of popular Twitter accounts with structured knowledge bases, including Wikidata\footnote{https://www.wikidata.org/} and DBpedia\footnote{https://www.dbpedia.org/}. which link entities to a hierarchy of semantic types. Prior work has mapped this information onto a fine-grained schema of semantic categories that are prevalent on social media~\cite{drukerman24}. Given this typing information, candidate entities of a desired type $\tau$ can be identified and ranked according to their social similarity to the user’s representation.

\section{Experiments: Inferring user preferences}
\label{sec:experiments}

We define the social recommendation task as follows. Given a user embedding $u_i$, we are interested in ranking candidate entities of interest ,of some specified semantic type $\tau$, by their relevance to the user. Our experiments implement a link prediction evaluation setup. We rank the candidate entities by their cosine similarity to the user profile, having first removed any of the candidate entities from the user's representation.
Entities that the user is known to follow are treated as relevant items, where it is desired that those entities be placed at the top of the ranked lists. To assess the generalization capabilities of social user modeling, we infer user preferences across 14 topical domains.

\subsection{Experimental dataset and setup}
\label{sec:social_network_data}

\begin{table*}[t]
\begin{footnotesize}
\begin{tabularx}{\textwidth}{l|X}
\textbf{Category} & \textbf{Target entities} \\
\hline
Musical artists	&	Justin Bieber, Katy Perry, Rihanna, Taylor Swift, Ariana Grande, Justin Timberlake, Selena Gomez, Britney Spears, Miley Cyrus, Wiz Khalifa, Kanye West, Eminem, Nicki Minaj , Snoop Dogg, John Legend, Kendrick Lamar, The Weeknd, Cher, Megan Thee Stallion, Bette Midler	\\
\hline
News	&	CNN Breaking News, The New York Times, SportsCenter, BBC News (World), The Economist, Reuters, Fox News, The Washington Post, ABC News, The Associated Press, HuffPost, The Onion, Guardian, Bleacher Report, NBC News, Usa Today, MSNBC, NY Post, Maddow blog, One America News	\\
\hline
Comedian	&	Ellen DeGeneres, Jimmy Fallon, Daniel Tosh, Stephen Colbert, Ricky Gervais, Sarah Silverman, Jimmy KImmel, Russell Brand, Bill Maher, Trevor Noah, Joe Rogan, Chris Rock, Seth Meyers, Amy Schumer, Jerry Seinfeld, Jim Gaffigan, David Spade, Billy Eichner, Brian Brushwood, Dave Chappelle	\\
\hline
Politicians	&	Barack Obama, Hillary Clinton, Michelle Obama, Bernie Sanders, Bill Clinton, Alexandria Ocasio-Cortez, Elizabeth Warren, Joe Biden, Nancy Pelosi, Kamala Harris, Chuck Schumer, Pete Buttigieg, Beto O'Rourke, Preet Bharara, Ted Lieu, Eric Swalwell, Stacey Abrams, Ron DeSantis	\\
\hline
TV station	&	CNN, ESPN, BBC News (World), Fox News, MTV, E! News, Discovery, Food Network, NBA TV, NFL Network, E! Entertainment, The Weather Channel, CNBC, MSNBC, NBC, CSPAN, Comedy Central, One America News, BlazeTV	\\
\hline
Actors	&	Jim Carrey, Tom Hanks, Robert Downey Jr, Chris Evans, Seth Rogen, Chris Pratt, Anna Kendrick, Chrishems Worth, Mark Ruffalo, Stevec Arell, John Cleese, Tom Holland, Patrick Stewart, Kumail Nanjiani, James Woods, Levarb Urton, Jordan Peele, Steven Crowder, Terrence K Williams, Ron Perlman	\\
\hline
TV show	&	The Daily Show, The Tonight Show, Good morning America, South Park, Stranger Things, NBA 2K, Saturday Night Live, Game Grumps, Sponge Bob, Mark R. Levin, Entertainment Tonight, World News Tonight, FOX \& friends, Maddow Blog, PBS News Hour, 60 Minutes, NBC Nightly News, Animalcro Ssing, Morning Joe	\\
\hline
Sports team	&	Golden State Warriors, Miami HEAT, FaZe Clan, Chicago Bulls, Boston Celtics, New York Yankees, Cleveland Cavaliers, Denver Broncos, Chicago Cubs, Green Bay Packers, San Francisco 49ers, Los Angeles Dodgers, New York Knicks, Red Sox, Chicago Bears, Las Vegas Raiders, Kansas City Chiefs, Atlanta Braves, New York Jets, New York Mets	\\
\hline
Fashion	&	H\&M, Louis Vuitton, Versace, La Coste, Under Armour, Balenciaga, Levi's, Hugo Boss, Nautica, Stussy, Fashion Nova, Juicy Couture, Ann Taylor, Tillys, Nine West, True Religion, Henri Bendel, prAna, Hanes, Legendary Whitetails	\\
\hline
Journalist	&	Ezra Klein, Chris Hayes, Joy Reid , Christopher Cuomo, Jim Acosta, Maggie Haberman, David Corn, David Fahrenthold, Daniel Dale, Katy Tur, Ari Melber, April Ryan, Robert Costa, Stephanie Ruhle, Philip Rucker, Kaitlan Collins, Aaron Rupa, Benny Johnson, Jonathan Swan, Elie Mystal	\\
\hline
TV host	&	Laura Ingraham, Lou Dobbs, Adam Richman, Buddy Valastro, Steve Inskeep, Guy Adami, Jeff Zeleny, Cari Champion, Jenna Compono, Abby Huntsman, Nicole Curtis, Mel Robbins, Clayton Morris, Ty Pennington, Kacie McDonnell, Gio Benitez, Pat Kiernan, Matt Gutman, Ziya Tong, Lulu Miller	\\
\hline
Film	&	Divergent, What Ted Said, Sense8, Star Trek, Disney's Frozen 2, Paddington, Toy Story 4, La La Land, Cowspiracy, The Hateful Eight, An Open Secret, Sausage Party, Demon House, Blade Runner, The Peanuts Movie, History Center, Hidden Figures	\\
\hline
Food chains	&	McDonalds, Chick-fil-A, Chipotle, KFC, What a burger, Dairy Queen, Buffalo Wild Wings, Applebees, Sonic Drive-In, Denny's Diner, Chilis, Steak 'n Shake, Raising Cane's, Jack in the Box, TGIFridays, Cracker Barrel, Hooters, White Castle, Texas Roadhouse, Zaxby's	\\
\hline
Car makers	&	Ford, Chevrolet, Jeep, Porsche, Lamborghini, Toyota, Mercedes Benz, Audi, Honda, BMW, Subaru, Rivian, Ferrari, Rolls-Royce, Jaguar, Kia, Bugatti, Volvo, Nissan	\\
\hline
\end{tabularx}
\caption{The target categories and entities in our experiments. Both categories and entities are listed in descending order of aggregate number of their followers (as of 2022).}
\label{tab:targets}
\end{footnotesize}
\end{table*}

To construct the experimental dataset, we first manually selected semantic categories that are likely to be of interest to a broad range of users, such as {\it musical artists}, {\it sports teams}, {\it TV shows}, and {\it politicians}. For each category, we identified relevant entities based on their semantic types obtained through alignment with DBpedia. We then selected the most popular entities per category, determined by their number of followers in Twitter. In total, our dataset includes 14 target categories, each with 20 candidate entities, resulting in 280 entities overall. Table~\ref{tab:targets} lists all categories and target entities. Both categories and entities are ordered in descending order of their cumulative number of Twitter followers (as of 2022). For example, entities in the {\it musical artists} category have the highest aggregate follower counts, and {\it Justin Bieber} is the most-followed entity within this category. In the following stage, for each candidate entity, we sampled random Twitter users who followed that entity, suggesting that it was of interest to them. In sampling, we referred to a large pool of 40K random user accounts, based in the U.S. Overall, we sampled 50 distinct users per entity. This procedure yielded a dataset of roughly 1K users associated with each target category, i.e. 14K users overall. 

In the experiments, we focus on each category in turn, ranking the candidate entities by their similarity to each user profile. Importantly, we remove any of the candidate entity embeddings from the composite social user embedding, simulating a link prediction setup in this fashion. While each user is known to follow at least one entity out of the candidate list, users may follow additional entities from the target category. Accordingly, we identified all of the target entities followed by each user as relevant responses. The average number of relevant responses per user in each category is detailed in Table~\ref{tab:rec_results}; e.g., the users selected per the {\it musical artists} category follow 7.6 entities from the candidate list on average. While other candidate entities may be of interest to the user, link prediction is commonly used as a  proxy for relevance judgments. Evaluating multiple methods using the same test set allows to compare their performances under the same conditions~\cite{pritskerIUI17}. 

\subsection{Results}

\begin{table}[t]
\begin{small}
\begin{tabular}{lc|c|cccc}
	&		Entities	&	Popularity	&	\multicolumn{4}{c}{Social user embeddings}	\\
Category	&	/ user	& (followers)	&	All	&	$\Delta$	& 	Sampled & $\Delta$ \\
\hline
Musical artists	&	7.6	&	0.520	&	0.633	&	22\%	&	0.591	&	14\%	\\
News outlets	&		8.2	&	0.581	&	0.692	&	19\%	&	0.661	&	14\%	\\
Comedians	&		6.8	&	0.545	&	0.599	&	10\%	&	0.594	&	9\%	\\
Politicians	&		11.2	&	0.720	&	0.777	&	8\%	&	0.755	&	5\%	\\
TV stations	&	4.5	&	0.469	&	0.618	&	32\%	&	0.603	&	29\%	\\
Actors	&	6.6	&	0.521	&	0.647	&	24\%	&	0.602	&	16\%	\\
TV shows	&	3.9	&	0.365	&	0.635	&	74\%	&	0.551	&	51\%	\\
Sports teams	&	4.3	&	0.349	&	0.510	&	46\%	&	0.385	&	10\%	\\
Fashion	&	2.6	&	0.372	&	0.447	&	20\%	&	0.373	&	0\%	\\
Journalists	&	10.1	&	0.622	&	0.675	&	9\%	&	0.640	&	3\%	\\
TV hosts	&	1.8	&	0.301	&	0.428	&	42\%	&	0.394	&	31\%	\\
Films	&	1.4	&	0.221	&	0.399	&	81\%	&	0.241	&	9\%	\\
Food chains	&	5.2	&	0.492	&	0.499	&	1\%	&	0.481	&	-2\%	\\
Car makers	&	5.7	&	0.472	&	0.437	&	-7\%	&	0.434	&	-8\%	\\
\hline		\multicolumn{2}{l|}{Overall results:} &	0.468	&	0.571	&	22\%	&	0.522	&	12\%	\\
\end{tabular}
\end{small}
\caption{Entity ranking results [MAP] using multiple approach: A popularity ranking, based on the number of followers; and, ranking by social similarity, representing users based on all Twitter entities (`'all'), or using the experimental entities only (`Sampled').}
\label{tab:rec_results}
\end{table}

Table~\ref{tab:rec_results} presents the ranking results in terms of mean average precision (MAP) per category, having the entities that each user followed considered as correct answers. To gauge the benefit of user modeling, we compare the results with a popularity baseline, having the candidates ranked based on the number of followers in Twitter. Popularity-based ranking serves as a strong non-personalized baseline in recommender systems~\cite{jiSIGIR20}. As shown in the table, ranking the entities by social similarity to the user yields performance gains for 13 out of 14 categories, where the relative gain in MAP exceeds 20\% in the majority of cases (8/14). In particular, high gains are achieved for the categories of {\it musical artists} (22\%), TV stations (32\%), {\it actors} (24\%) {\it news outlets} (46.9\%), TV shows (74\%), sports teams (46\%), {\it fashion} (20\%), {\it TV hosts} (42\%) and {\it movies} (81\%). It is interesting to note that for the categories of {\it TV hosts} and {\it movies} there is a small number of accounts that users follow on average (1.4-1.8), where this is reflected in relatively low MAP results. Also, while users who follow {\it food chains} and {\it car makers} tend to co-follow multiple entities in these domains (5.2-5.7), there are fewer users overall who are interested who follow entities in these categories in Twitter. Perhaps due to this reason, cross-domain inference towards these domains is more challenging. Nevertheless, as shown in Table~\ref{tab:rec_results}, the average improvements over popularity ranking across all categories in the experimental dataset is as high as 22\%. We therefore conclude that modeling users in a social embedding space enables the prediction of their personalized preferences in a variety of topics. 

\paragraph{Closed-world evaluation.} 
Table~\ref{tab:rec_results} reports also the results of another experiment, in which we constructed the social user representations only from relevant target entities within our experimental dataset. In other words, the social embedding of each user is composed in this case based on the target entities that they follow in all (13) domains, except for the target domain. This `closed world' setup enables a restrictive evaluation, limiting exposure to relevant information about other entities in the target domain; for example, in ranking candidate {\it musical artists}, this setup is intended to eliminate any other musical artists from the user's representation. As shown in Table~\ref{tab:rec_results}, also in this restrictive evaluation mode (`Sampled'), social modeling yields substantial performance gains, resulting in 12\% improvement in MAP on average compared to the popularity method. These results therefore confirm that social user modeling enables the prediction of personal preferences across domains. We note that the lower gains observed in this setup are also due to limiting the social information that is modeled per user--whereas users in our datasets generally follow 1,972 entities on average  across any domain, in the `closed world' setup, the user representation is based on the embeddings of only 40 entities on average, spread over 13 categories.

\subsection{Analysis: How much data is needed to achieve effective personalization?}

\begin{figure*}[t]
\centering
\includegraphics[width=5.2cm]{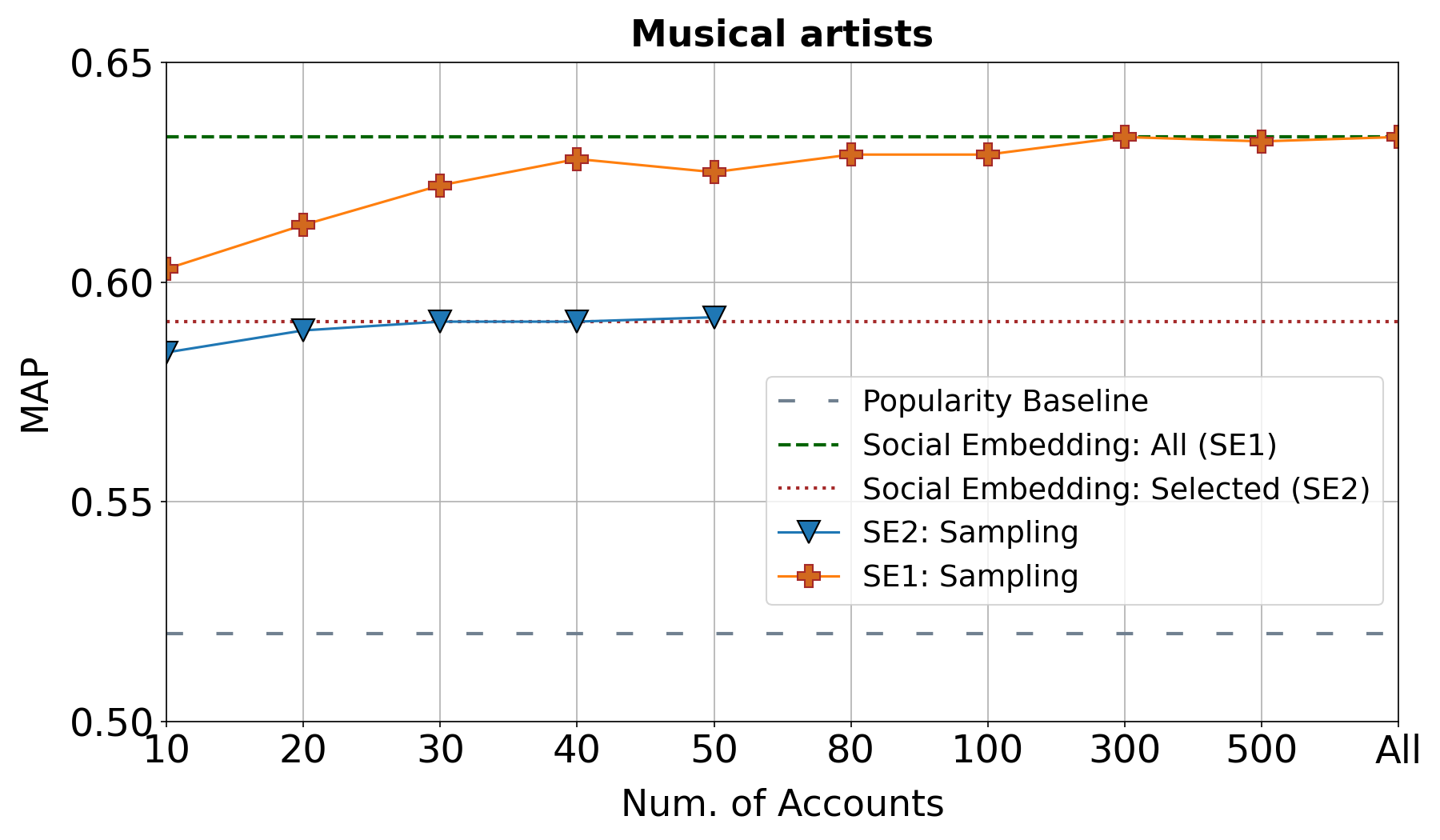}
\includegraphics[width=5.2cm]{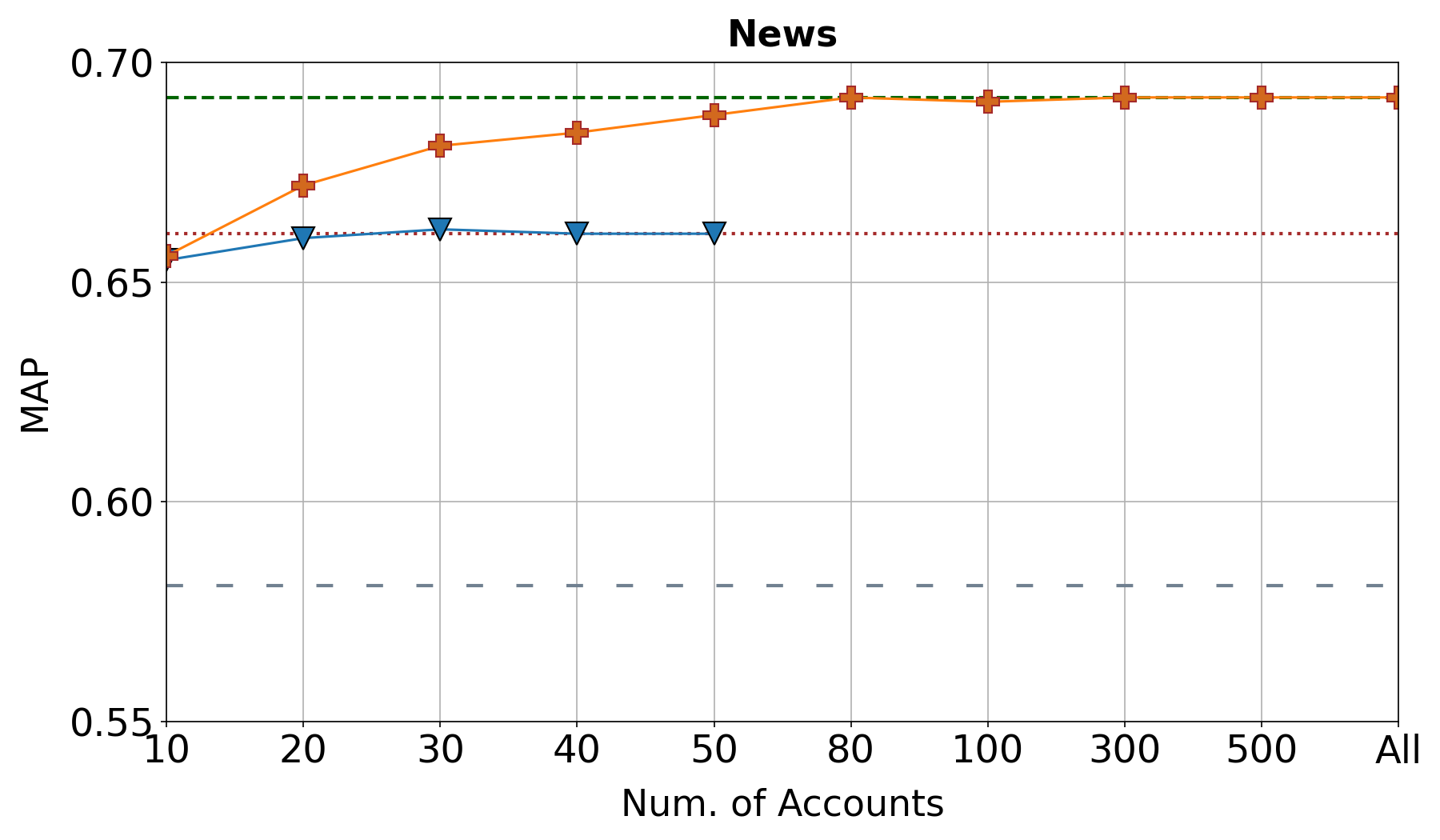}
\includegraphics[width=5.2cm] {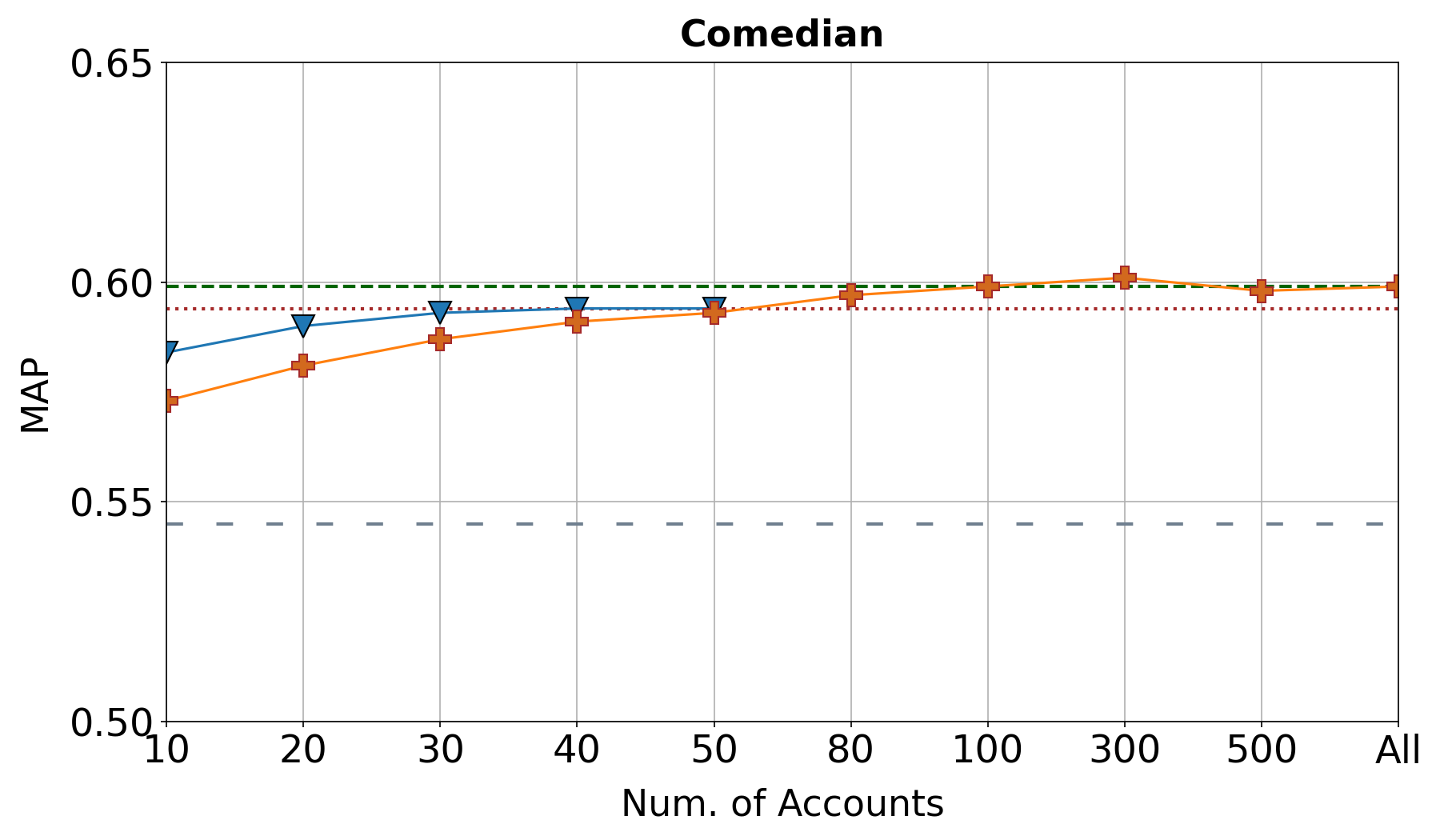}
\\
\vspace{0.3cm}
\includegraphics[width=5.2cm]{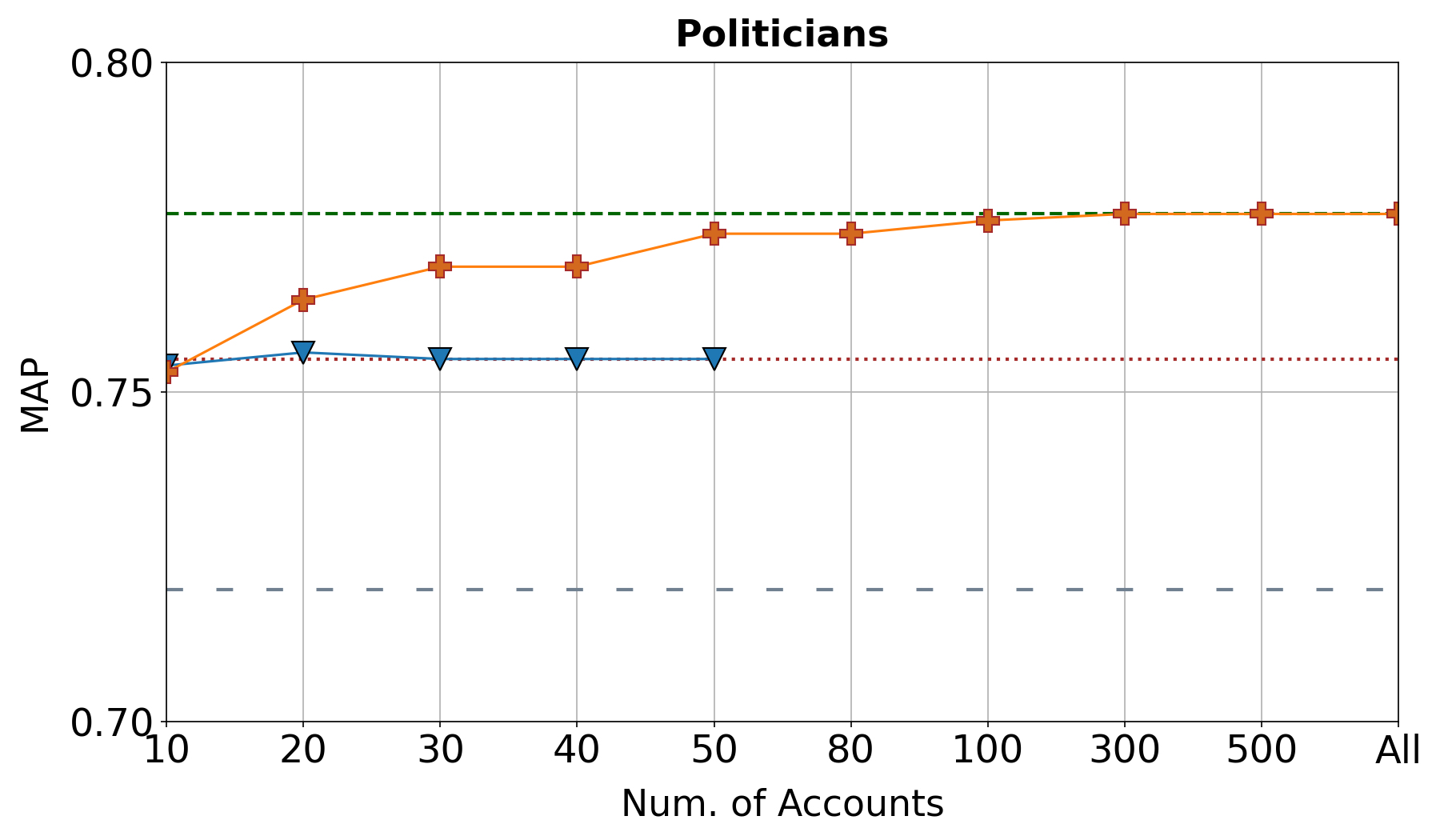}
\includegraphics[width=5.2cm]{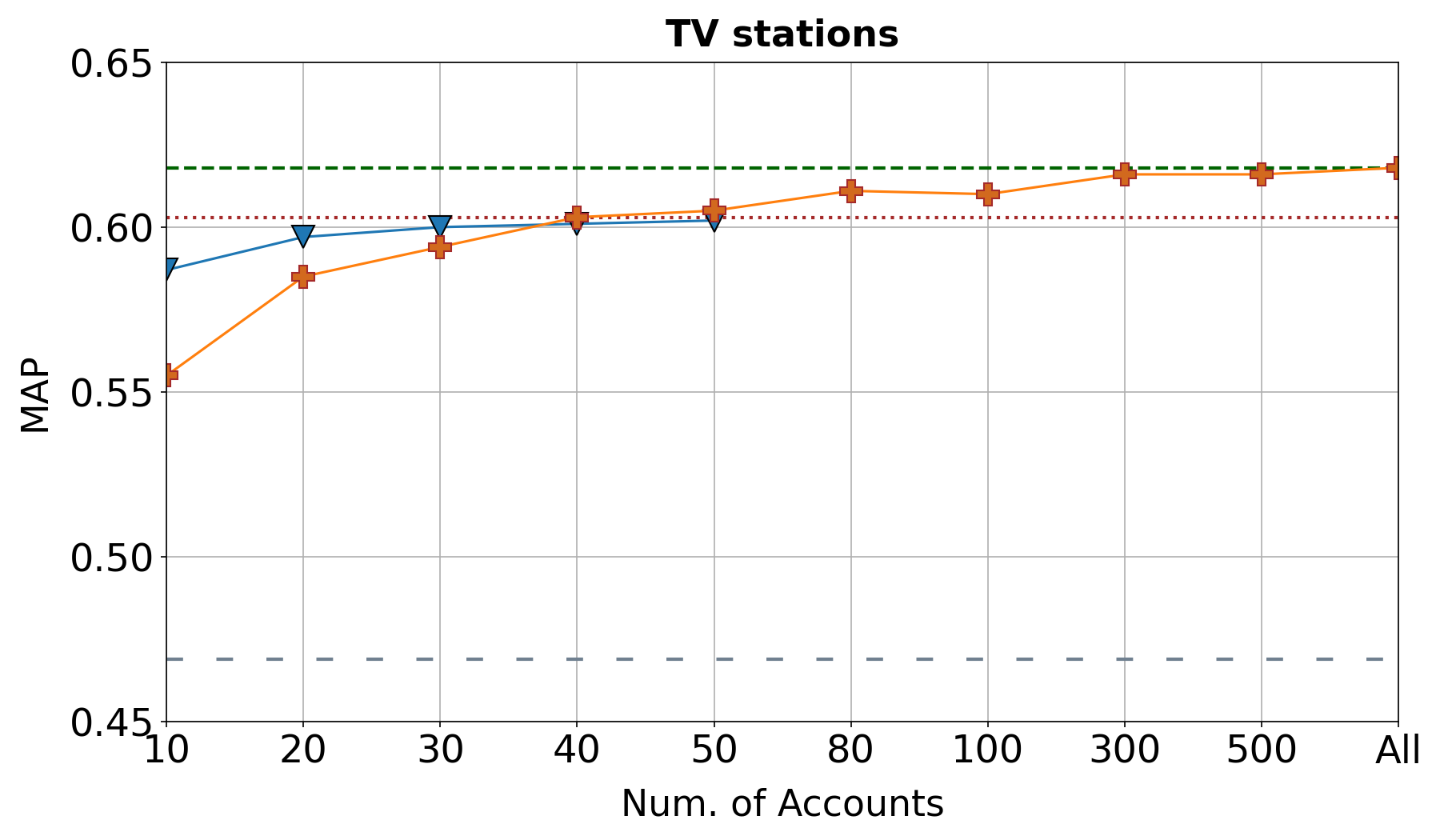}
\includegraphics[width=5.2cm] {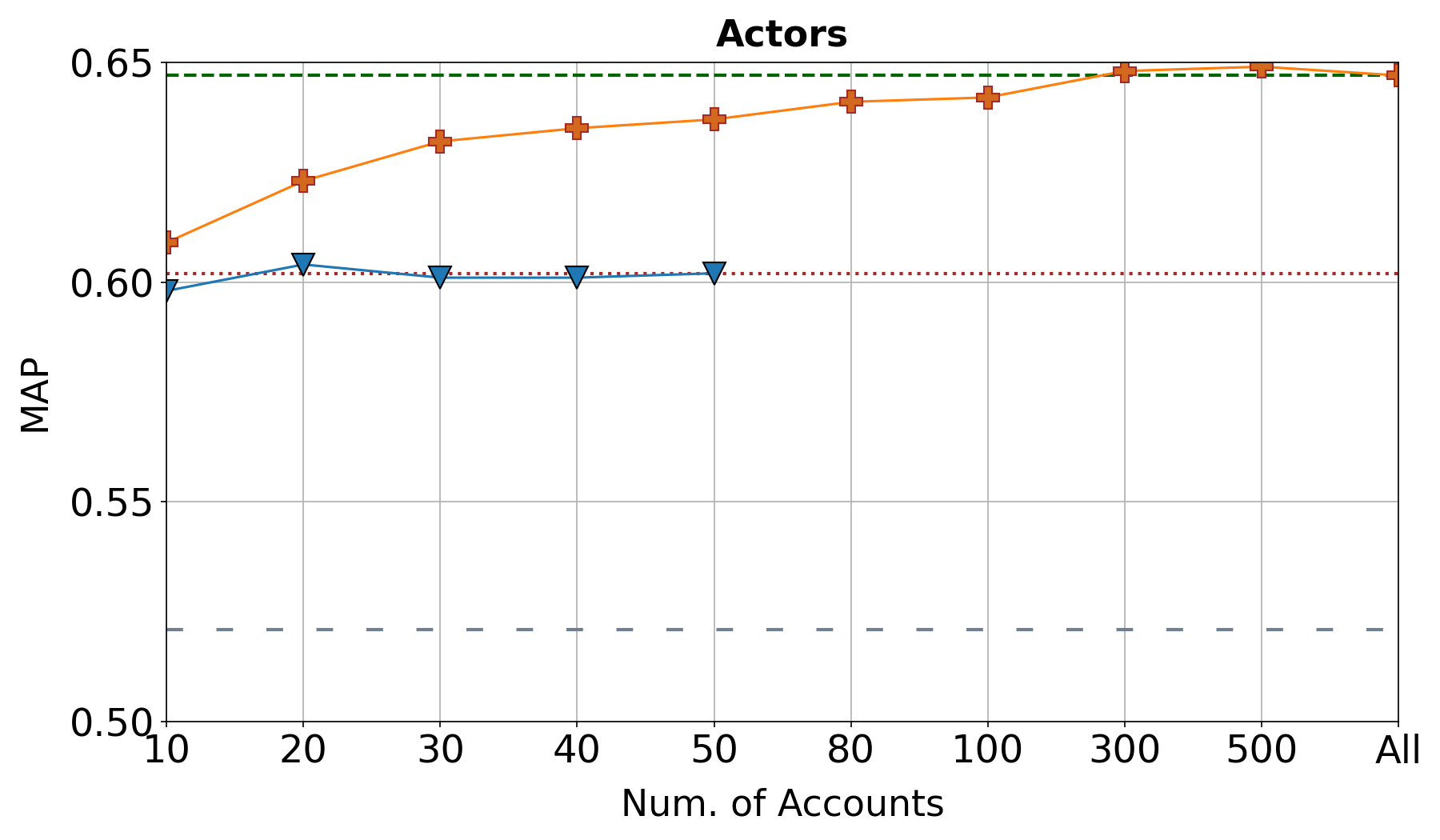}
\\
\vspace{0.3cm}
\includegraphics[width=5.2cm]{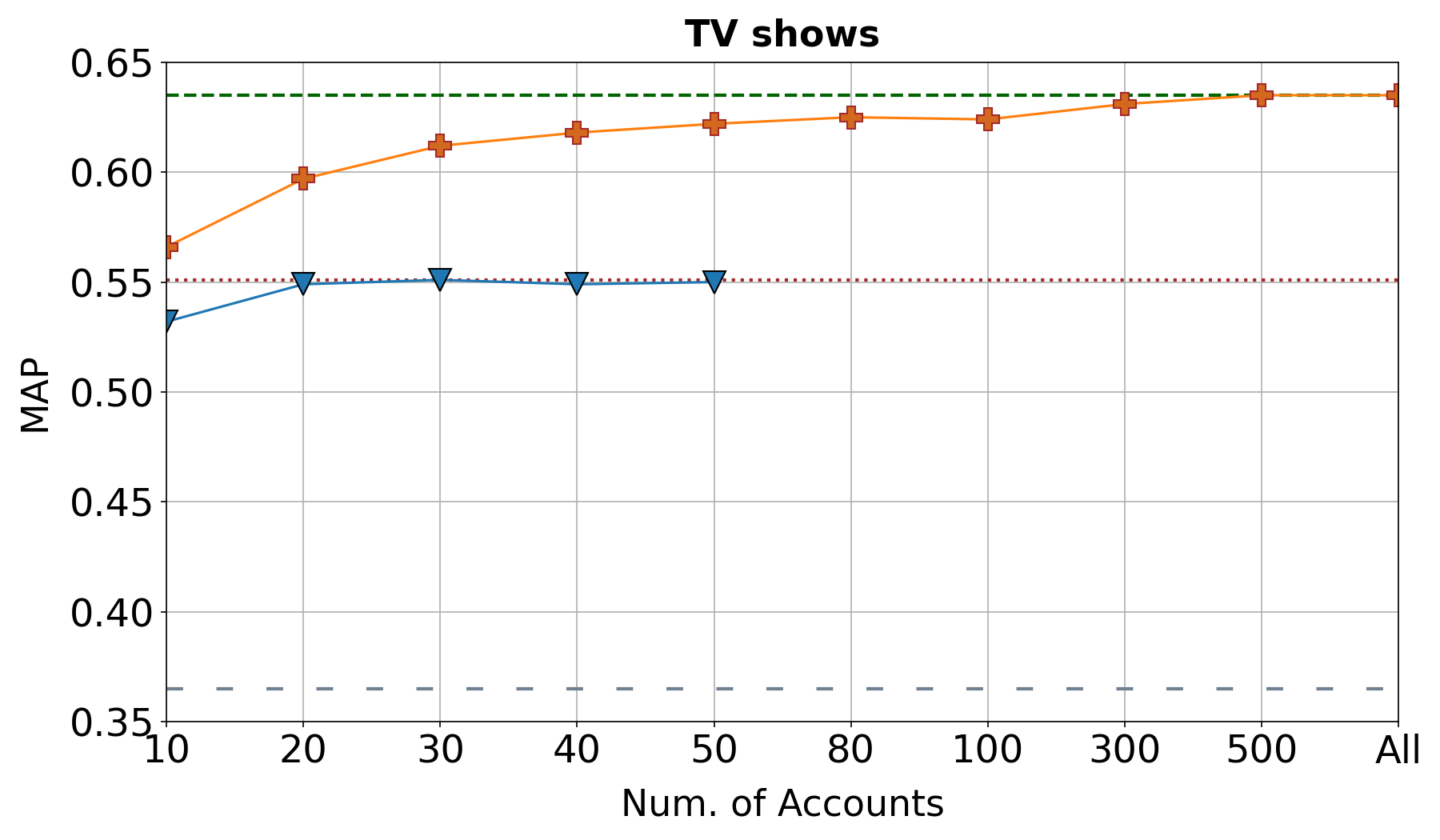}
\includegraphics[width=5.2cm]{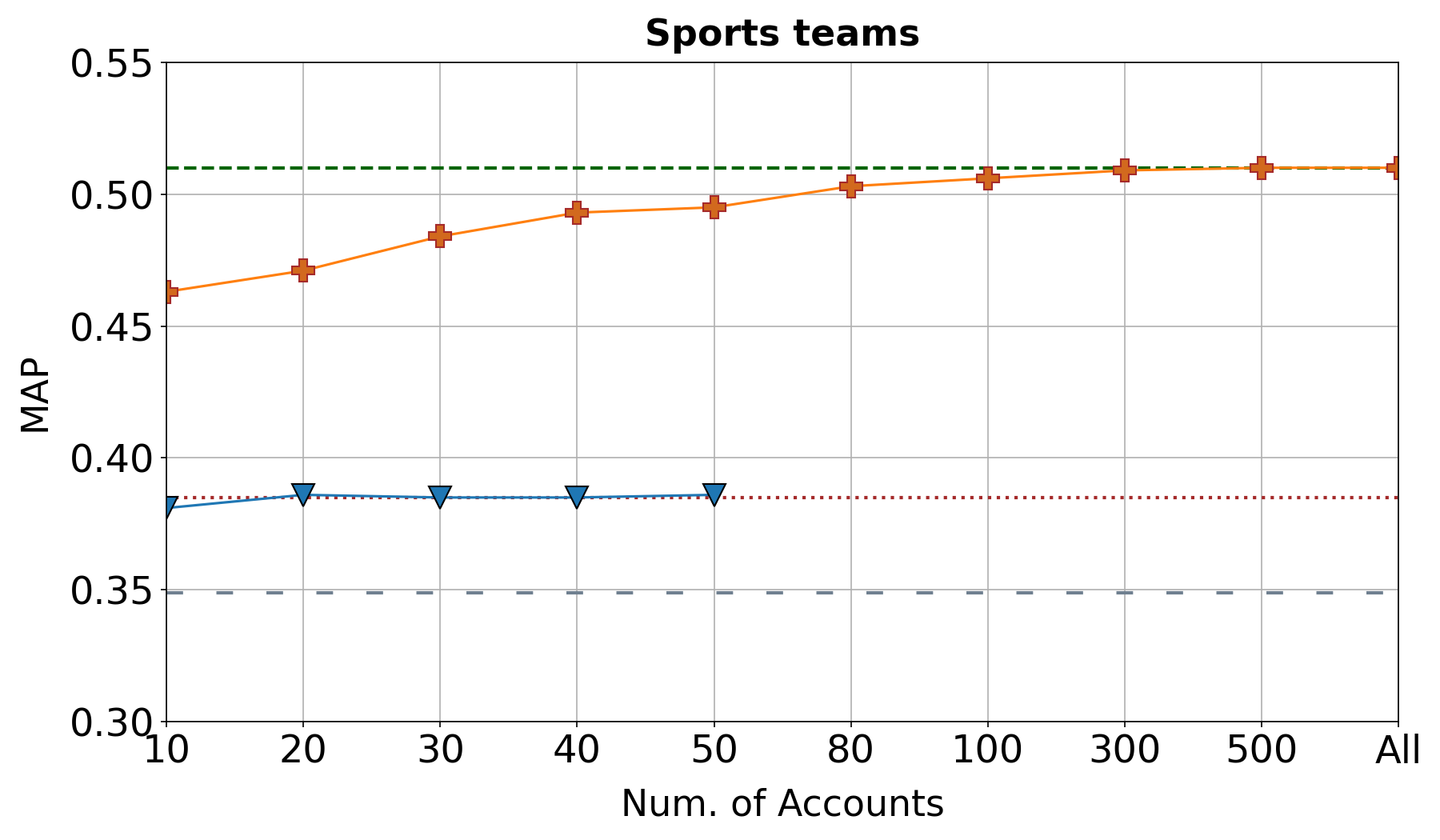}
\includegraphics[width=5.2cm] {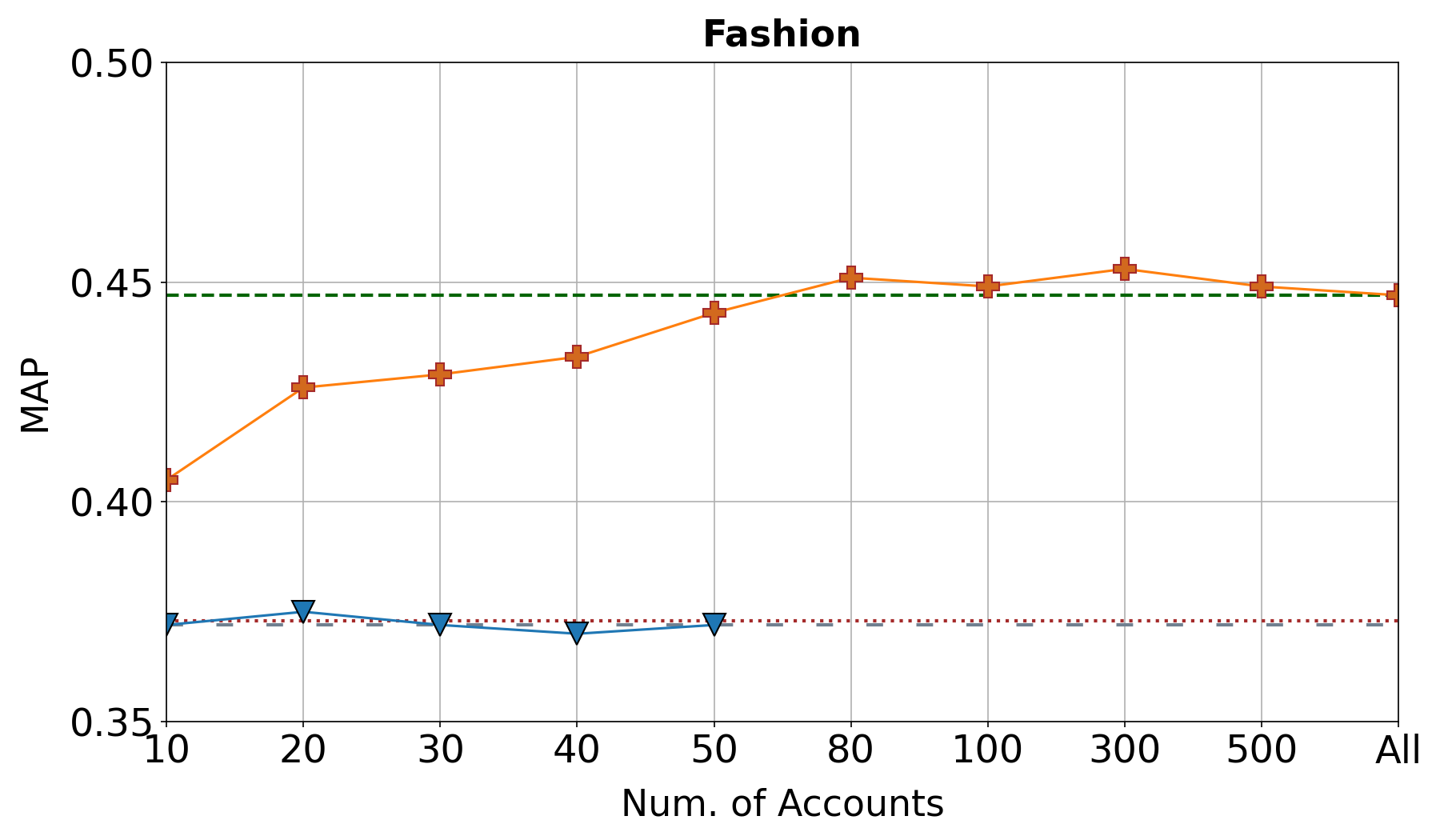}
\\
\vspace{0.3cm}
\includegraphics[width=5.2cm]{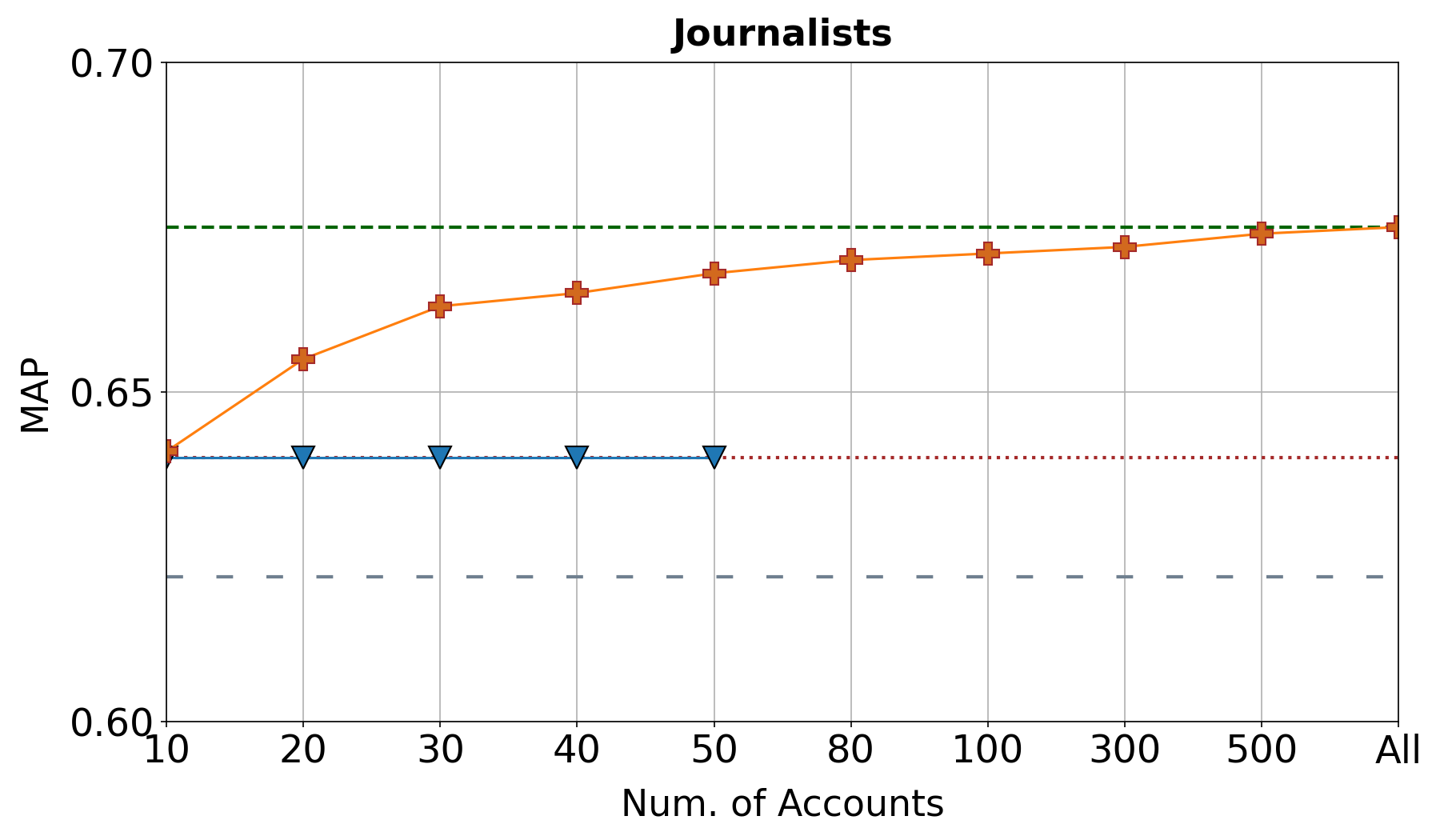}
\includegraphics[width=5.2cm]{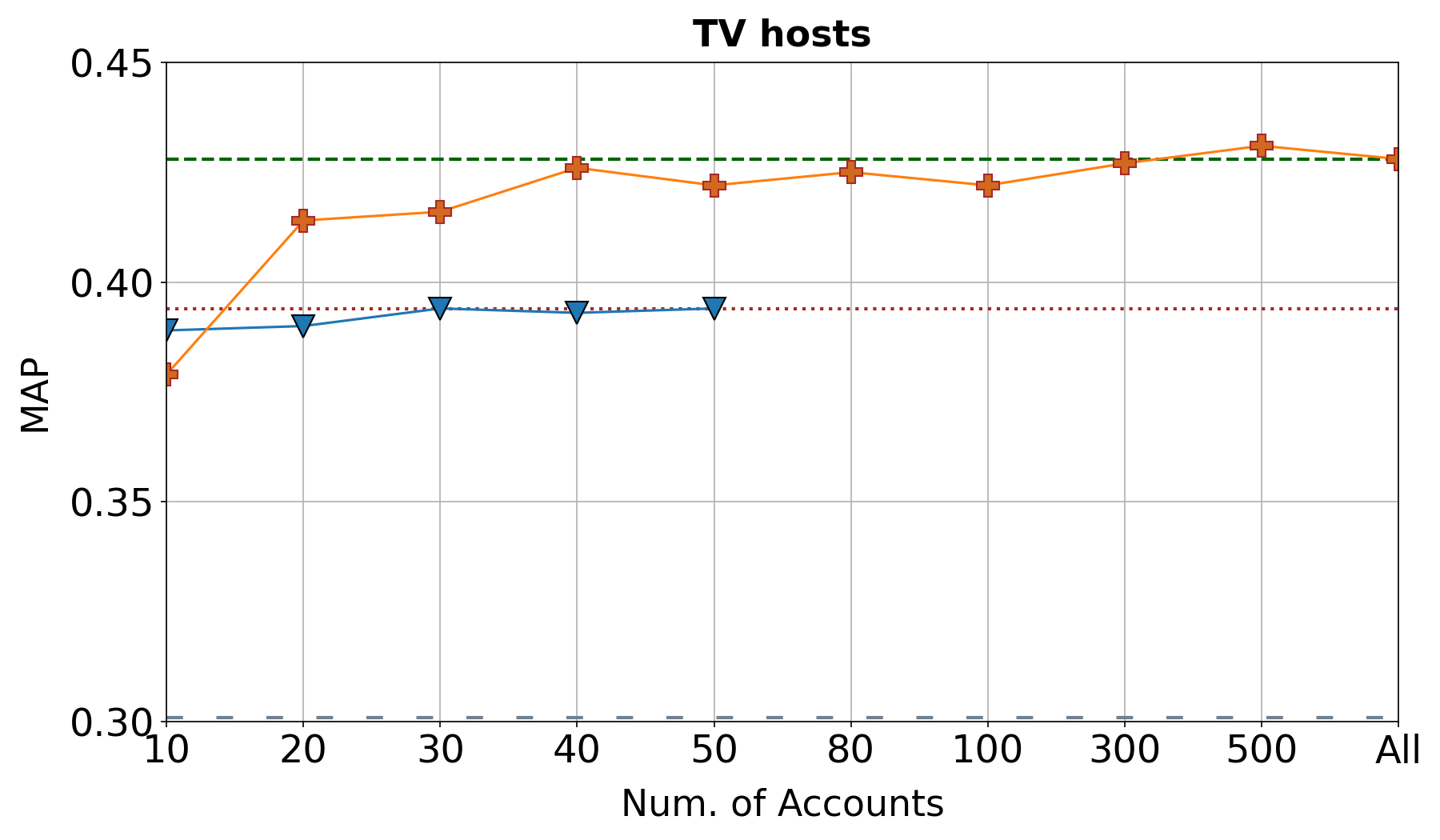}
\includegraphics[width=5.2cm] {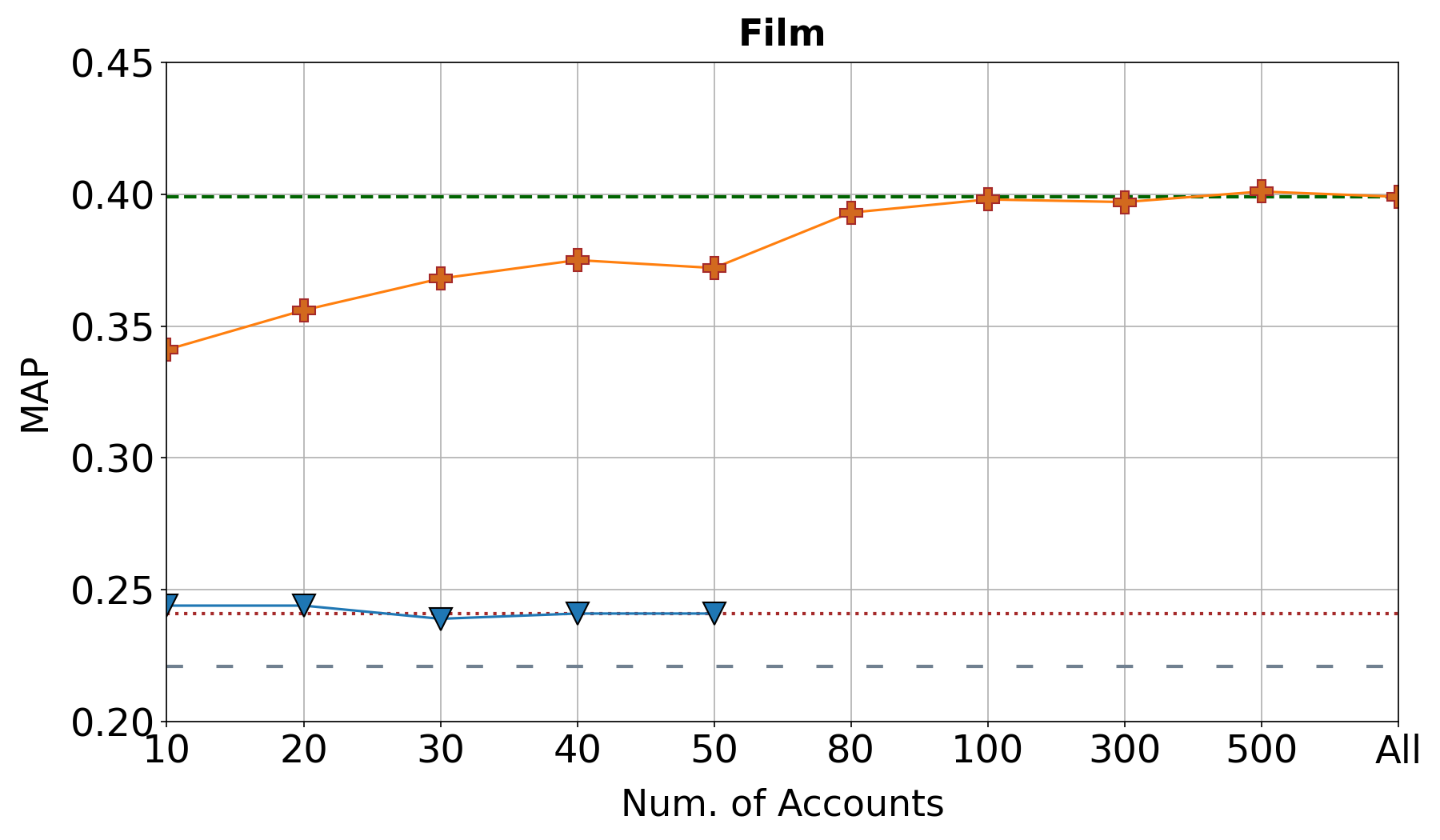}
\\
\vspace{0.3cm}
\includegraphics[width=5.2cm]{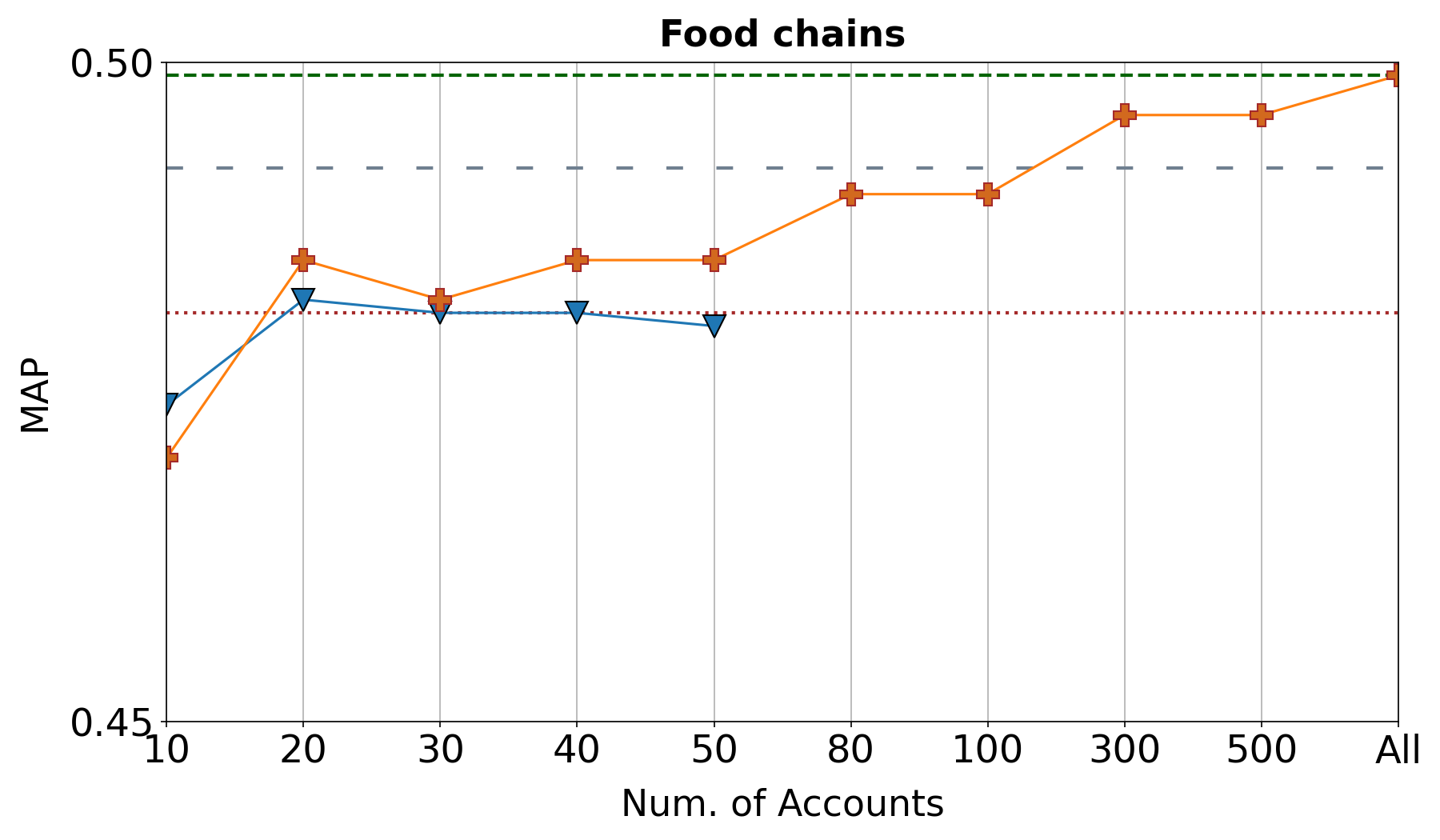}
\includegraphics[width=5.2cm]{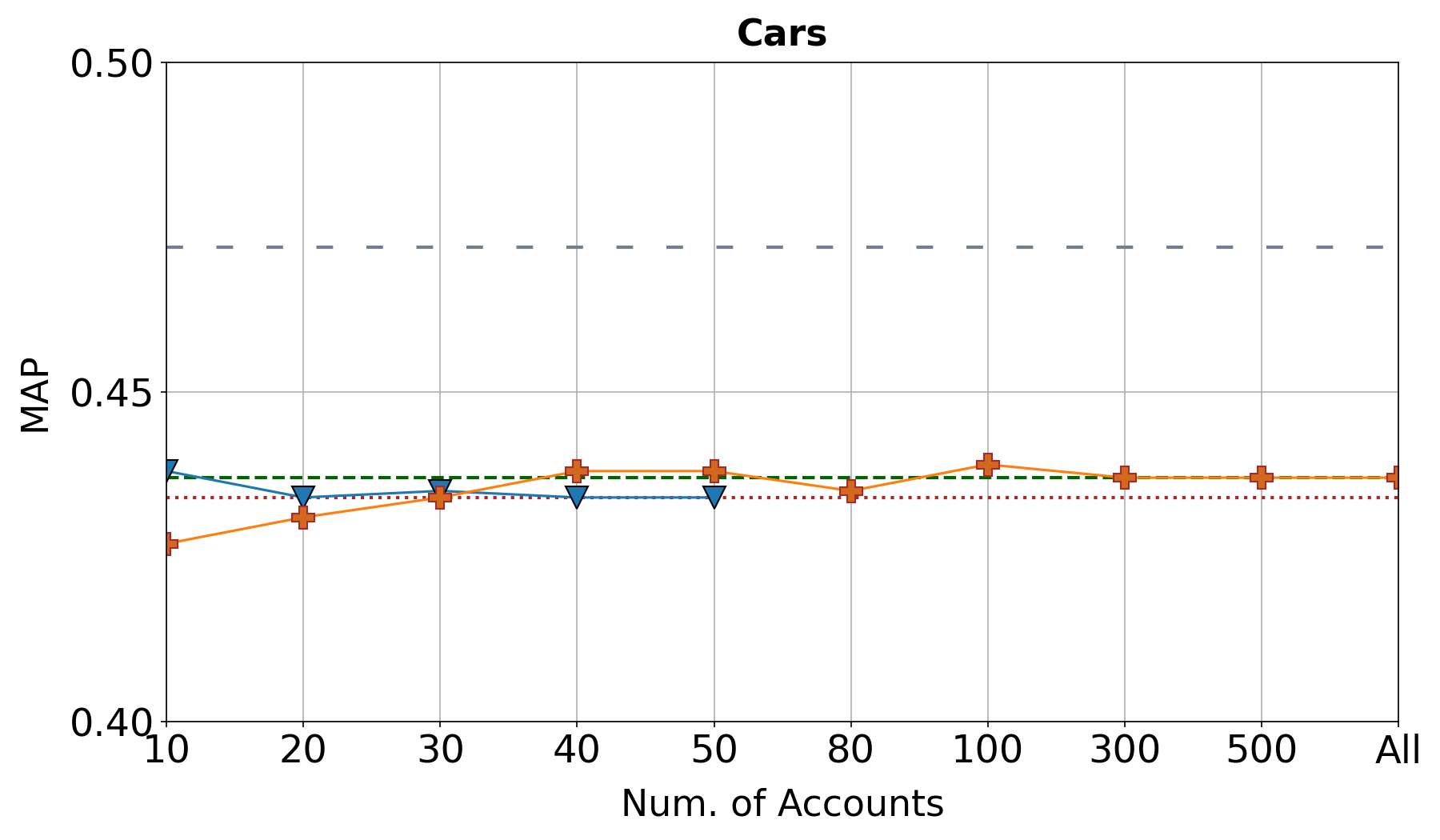}
\\
\caption{MAP results: Varying the number of entities that comprise the social user representations. The entities are either sampled from all entities (SE1) or a `closed-world' selection (SE2). The horizontal lines depict the reference results in Table~\ref{tab:rec_results}.}
\label{fig:vary}
\end{figure*}

A following question is, {\it how many entities comprise a high-quality user representation, which supports effective personalization?} To address this question, we report another set of controlled experiments, in which we sample the entities that the user representations are composed from, limiting the number of sampled entities to some specified size $k$. Let us first consider the unrestricted evaluation setup. Figure~\ref{fig:vary} reports MAP results on our test set for varying values of $k$ (`SE1:sampling'), starting from as few as $k=10$ entities, and up to $k=500$. The results show that preference prediction performance improves as increasing number of entities are utilized for social user modeling. The improvements subside however beyond $k=100$. As illustrated in the figure, the results using the sampled entities converge quickly with our results using the full user profiles ('SE1:All'). $k=50$ yields 98.4\% of the average MAP score reported in our full experiments (0.562 vs. 0.571). In fact, as few as $k=10$ or $k=20$ yield 93.1\% (0.532) and 96.0\% (0.548) of the average best MAP scores, respectively. Figure~\ref{fig:vary} reports similar trends for the `closed world' experiments (`SE2: Sampling'). In this setup, there are fewer entities associated with each user, hence the number of sampled accounts per user reaches $k=50$ at most. In this case, performance converges faster, reaching 99.0\% of the performance using all of the relevant entities ('SE2: All')  using $k=10$ (0.517 vs. 0.522), and 99.9\%--using $k=20$ (0.521). 
This analysis suggests that our approach of social user modeling is robust to sparse information, enabling effective personalization given limited inputs. Later, we discuss the implications of these findings for a cold-start scenario, where users may be asked to indicate a small number of entities that they favor during onboarding.

\section{The socio-demographics underlying social user modeling}
\label{sec:socio}

\begin{figure*}[t]
\centering
\includegraphics[width=5cm]{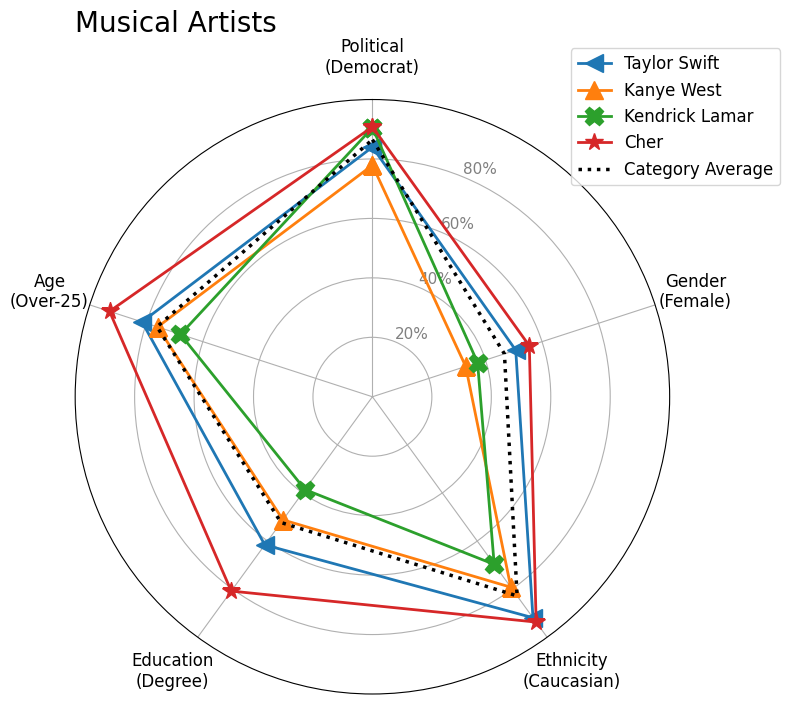}
\includegraphics[width=5cm]{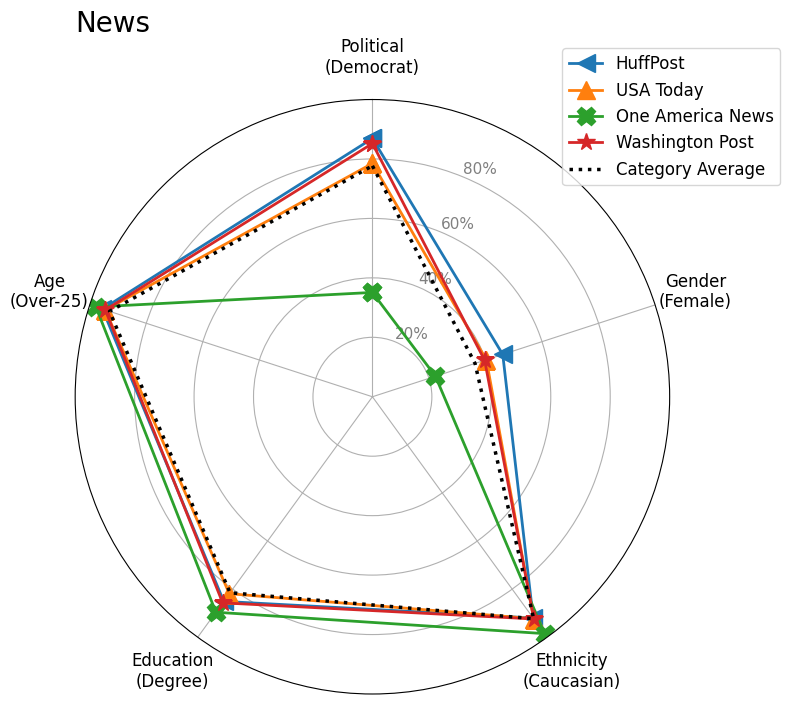}
\includegraphics[width=5cm] {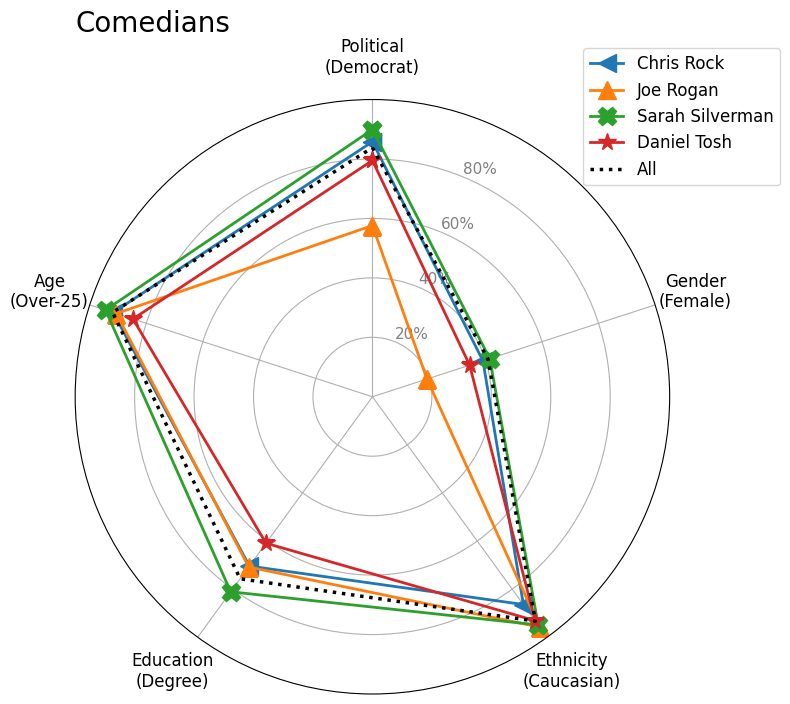}
\\
\vspace{0.3cm}
\includegraphics[width=5.2cm]{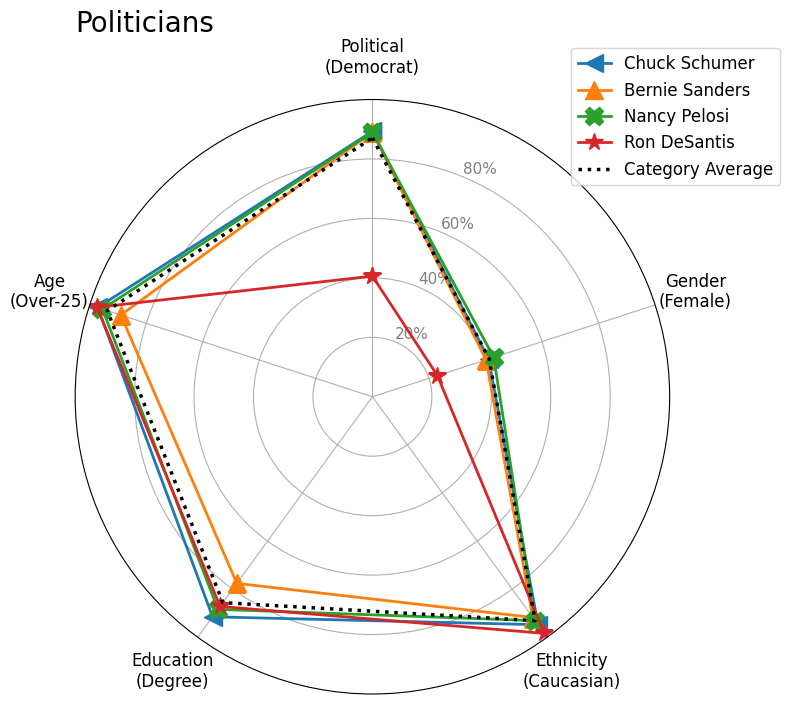}
\includegraphics[width=5.2cm]{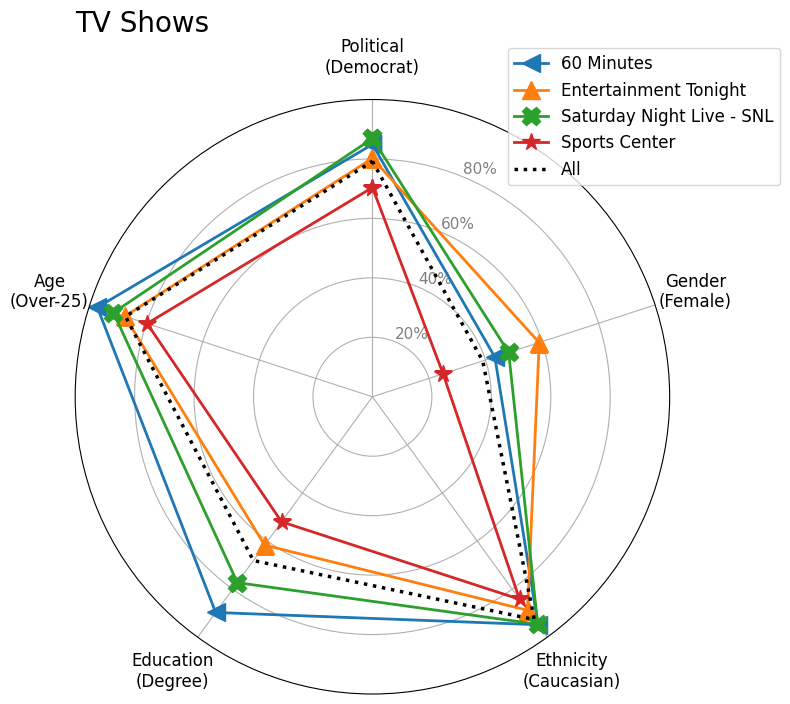}
\includegraphics[width=5.2cm] {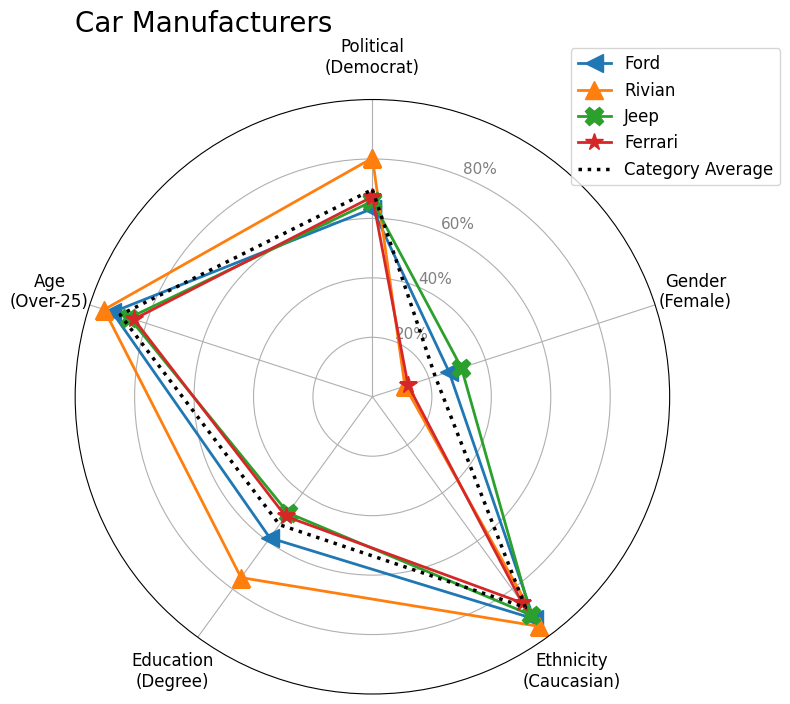}
\\
\caption{Aggregate socio-demographic profiles of the users who follow selected categories and entities in our experimental dataset; see an explanation and discussion in Sec.~\ref{sec:socio}.}
\label{fig:radar_plots}
\end{figure*}

It has been shown that a variety of personal traits can be inferred with high precision based on the accounts that one chooses to follow on social media~\cite{culotta2015predicting,dangur20,muellerCSCW21}. We exploit classifiers that were trained in a supervised fashion to predict the users' socio-demographic traits from the social network embedding vectors~\cite{lotanPLOS23}. These classifiers rely on human-labeled datasets, annotated using binary distinctions of {\it gender}, {\it age}, {\it ethnicity}, {\it education level}, and {\it political affiliation}~\cite{volkovaACL16}. By applying these classifiers to the users included in our experimental dataset, we characterize the population of users who follow target entities in multiple domains. A previous study showed that sociodemographic factors underlie individuals’ stances and life choices~\cite{pendzelEMNLP25}. The analysis presented in this work further demonstrates that the latent dimensions encoded in social user representations enable the generalization of personal preferences across different topical domains.

Figure~\ref{fig:radar_plots} presents an aggregate view of the resulting statistics for selected categories and entities. The results are illustrated using radar plots. In these plots, each trait is represented using a dedicated axis, where the binary trait values reside on the circle's center and diameter, and the proportion of trait values within the relevant population of users is denoted using a point along that axis. Hence, the closed shape connecting the observed proportion markers over all traits represents a collective social profile. The profiles of users who follow each target entity are presented on top of each other, illustrating the social differences between these populations. For each category, the plot includes a dashed line which denotes the average socio-demographic statistics per all entities in that category. The aggregate statistics vary by category; for example, users who follow {\it politicians} tend to be older on average than the users who follow {\it musical artists} in our dataset. 

Let us take a closer look at the {\it politicians} category. We observe that the profiles of users who follow Chuck Schumer, Bernie Sanders and Nancy Pelocy are rather similar, with almost 90\% of their followers being classified as Democrats. Among these politicians, the followers of Bernie Sanders are somewhat younger, with a lesser proportion of academic education. In contrast, the users who follow Ron DeSantis are mostly Republican, dominated by men, and include the highest proportion of Caucasian white users. As another example, let us consider the category of {\it TV shows}. Among the presented entities, `E! Tonight' has the highest ratio of women among its followers, whereas `Sports Center' has the highest ratio of men; the `60 Minute' show has a notably high proportion of users who are over 25 years of age and hold academic degrees; and, the Saturday Night Live show has the highest proportion of followers identified as liberal. 

Since social user embeddings encode socio-demographic trait information, and given that these traits correlate with users' preferences of entities, social modeling enables the generalization of personal preference towards new entities and topical domains. Beyond coarse socio-demographic attributes, embeddings learned from large-scale social network data capture fine-grained latent dimensions related to hobbies and lifestyles, occupations, geopolitical contexts, and other aspects of social identity. Modeling these rich and intricate social contexts is therefore likely to be a key ingredient for effective cross-domain personalization.

\section{Implications: Towards cold-start social user modeling using LLMs}
\label{sec:llm}

So far, we have utilized social entity embeddings learned from a large sample of the Twitter network. Presumably, large language models (LLMs) also encode rich social information about entities, as they are trained on vast amounts of text spanning multiple genres, including data that captures co-occurrences of entity mentions. Since LLMs operate in a lexico-semantic embedding space, social user information can be conveyed to the model in the form of a list of entity names that the user is known to favor. Importantly, we do not assume that end users of conversational systems are members of social networks, nor that they are willing to disclose such information. We show that under such cold-start conditions, indicating only a small number of liked entities is sufficient for the LLM to effectively model users’ personal preferences.

We experimented with this approach using the conversation agent of OpenAI GPT-4o\footnote{Version 2025-01-01-preview}. A popularity ranking of the candidate entities in each category was elicited using the  prompt: "{\it Given the candidate entities specified below, produce a fully ordered recommendation list for a Twitter user to follow.}" Personalized rankings were generated using the prompt: "{\it Given a sampled list of accounts that a Twitter user follows, and a list of candidate entities, generate a prioritized list of entities that would like to follow.}" Table~\ref{tab:llm_results} shows MAP results, specifying $k=12,25$ entities per user, sampled among the experimental entities (`closed world' setting). We also report performance using $k=50$ entities, randomly sampled from the users' profiles. It is shown that the non-personalized rankings (`baseline') produced by the LLM yield comparable or better MAP results compared with Twitter popularity rankings (0.49 vs. 0.47). A main finding of this experiment is that modeling users as a list of popular entities that they like is highly effective, with as few as $k=12$ entities yielding 13\$ gains in MAP performance. If $k=50$ entities are specified per user, than performance gains rise to 23\%, 

\begin{table}[t]
\begin{small}
\begin{tabular}{lcc|cccc}
Category	&	Followers	&	Social	&	Baseline	&	$k=12$	&	$k=25$	&	$k=50$	\\
\hline
Musical artists	&	0.520	&	0.625	&	0.508	&	0.582	&	0.611	&	0.667	\\
News outlets	&	0.581	&	0.688	&	0.648	&	0.695	&	0.698	&	0.716	\\
Comedians	&	0.545	&	0.593	&	0.576	&	0.638	&	0.643	&	0.655	\\
Politicians	&	0.72	&	0.774	&	0.785	&	0.797	&	0.801	&	0.811	\\
TV stations	&	0.469	&	0.605	&	0.507	&	0.636	&	0.640	&	0.664	\\
Actors	&	0.521	&	0.637	&	0.511	&	0.605	&	0.603	&	0.662	\\
TV shows	&	0.365	&	0.622	&	0.427	&	0.595	&	0.612	&	0.637	\\
Sports teams	&	0.349	&	0.495	&	0.347	&	0.393	&	0.391	&	0.571	\\
Fashion	&	0.372	&	0.443	&	0.377	&	0.409	&	0.409	&	0.493	\\
Journalists	&	0.622	&	0.668	&	0.68	&	0.673	&	0.675	&	0.704	\\
TV hosts	&	0.301	&	0.422	&	0.313	&	0.436	&	0.451	&	0.522	\\
Films	&	0.221	&	0.372	&	0.216	&	0.306	&	0.308	&	0.399	\\
Food chains	&	0.492	&	0.485	&	0.492	&	0.496	&	0.504	&	0.496	\\
Car makers	&	0.472	&	0.438	&	0.472	&	0.480	&	0.487	&	0.409	\\
\hline
Average	&	0.468	&	0.562	&	0.490	&	0.553	&	0.560	&	0.600	\\
$\Delta$ LLM & & & & 13\% & 14\% & 23\% \\
\end{tabular}
\end{small}
\caption{LLM Results: A non-personalized baseline vs. personalized rankings generated given a set of entities of size $k$ per user. The leftmost columns include results from Table~\ref{tab:rec_results}.}
\label{tab:llm_results}
\end{table}

\begin{table}[t]
\begin{small}
\begin{tabular}{lccccccc}
 \hline
k/n	&	1	&	2	&	3	&	4	&	5	&	6	&	7	\\
    \hline
1	&	0.471	&	0.478	&	0.482	&	0.485	&	0.485	&	0.485	&	0.485	\\
2	&	0.476	&	0.490	&	\cellcolor{gray!50} 0.496	&	\cellcolor{gray!50} 0.500	&	\cellcolor{gray!50} 0.501	&	\cellcolor{gray!50} 0.501	&	\cellcolor{gray!50} 0.502	\\
3	&	0.476	&	0.493	&	\cellcolor{gray!50} 0.503	&	\cellcolor{yellow!50} 0.506	&	\cellcolor{yellow!50} \cellcolor{yellow!50} 0.509	&	\cellcolor{yellow!50} 0.511	&	\cellcolor{yellow!50} 0.511	\\
4	&	0.477	&	0.493	&	\cellcolor{gray!50} 0.500	&	\cellcolor{yellow!50} 0.508	&	\cellcolor{yellow!50} 0.511	&	\cellcolor{orange!50} 0.516	&	\cellcolor{orange!50} 0.516	\\
5	&	0.477	&	\cellcolor{gray!50} 0.495	&	\cellcolor{yellow!50} 0.507	&	\cellcolor{yellow!50} 0.512	&	\cellcolor{orange!50} 0.515	&	\cellcolor{orange!50} 0.517	&	\cellcolor{orange!50} 0.518	\\
\hline
\end{tabular}
\end{small}
\label{tab:cold}
\caption{MAP results for varying number of accounts ($n$=1-7) sampled per a varying number of categories ($k$=1-5) per user.}
\end{table}

In practice, we envision that user information can be collected through a friendly interface that serves as an engaging entry point to the conversational system. This can be achieved by first presenting users with a set of high-level interest categories (e.g., {\it politics}, {\it sports}, or {\it music}). For each selected category, the user may then be shown a curated list of popular entities and asked to indicate those that best match their personal tastes. Table~\ref{tab:llm_results} depicts ranking results by sampling the entities in this fashion, varying the number of categories (1-5) and number of entities per category (1-7). The table forms a `heat map', with colors noting increasing levels of performance per these parameter combinations. Overall, we find that these results justify the selection of 3-5 categories with 4-5 items per category by end users. 

\section{Conclusion}

We have shown that modeling users within a social embedding space based on the entities that they follow enables effective prediction of their preferences across topical domains. Our analyses placed emphasis on `cold start' conditions, showing that effective personalizaion is achieved also when social user representations are based on a small number of relevant entities. Importantly, we obtained consistent results using a conversation agent, suggesting that entity-based user modeling carries social information that is useful for cross-domain personalization also using the paradigm of LLMs. We will make our experimental dataset available to researchers. Future research may gauge the modeling of user preferences with respect to entities that are local or niche rather than world-known. The social entity embeddings may provide complementary coverage to LLM knowledge in such cases. We are further interested in assessing personalization on tasks such as personal search or text generation, and in integrating social user representation as contextual modality for LLMs.   

\section{Ethical statement}

This study was approved by the IRB (XXX/2X). The data were collected from Twitter for research purposes, and the user embeddings we use are effectively anonymized, as they are low-dimensional projections derived from aggregated follow information. Given the aggregation, obfuscation, and scale of the data, user identities cannot be recovered. Our analysis suggests that social user representations might encode social biases and stereotypes. We hope this work will motivate further research on understanding and mitigating such biases. Importantly, any use of personal information requires informing users of these risks and obtaining their consent.

\bibliographystyle{ACM-Reference-Format}
\bibliography{main}

@article{lotanPLOS23,
  title={Social world knowledge: Modeling and applications},
  author={Lotan, Nir and Minkov, Einat},
  journal={Plos one},
  volume={18},
  number={7},
  year={2023},
  publisher={Public Library of Science San Francisco, CA USA}
}

@inproceedings{personallmICLR25,
 author = {Zollo, Thomas and Siah, Andrew and Ye, Naimeng and Li, Li and Namkoong, Hongseok},
 booktitle = {International Conference on Representation Learning},
 title = {PersonalLLM: Tailoring LLMs to Individual Preferences},
 url = {https://proceedings.iclr.cc/paper_files/paper/2025/file/a730abbcd6cf4a371ca9545db5922442-Paper-Conference.pdf},
 year = {2025}
}

@inproceedings{transe,
author = {Bordes, Antoine and Usunier, Nicolas and Garcia-Dur\'{a}n, Alberto and Weston, Jason and Yakhnenko, Oksana},
title = {Translating embeddings for modeling multi-relational data},
year = {2013},
booktitle = {Proceedings of the 27th International Conference on Neural Information Processing Systems - Volume 2},
}

@inproceedings{shenACL18,
    title = "Baseline Needs More Love: On Simple Word-Embedding-Based Models and Associated Pooling Mechanisms",
    author = "Shen, Dinghan  and
      Wang, Guoyin  and
      Wang, Wenlin  and
      Min, Martin Renqiang  and
      Su, Qinliang  and
      Zhang, Yizhe  and
      Li, Chunyuan  and
      Henao, Ricardo  and
      Carin, Lawrence",
    booktitle = "Proceedings of the Annual Meeting of the Association for Computational Linguistics",
    year = "2018"
}

@inproceedings{twhinKDD22,
author = {El-Kishky, Ahmed and Markovich, Thomas and Park, Serim and Verma, Chetan and Kim, Baekjin and Eskander, Ramy and Malkov, Yury and Portman, Frank and Samaniego, Sof\'{\i}a and Xiao, Ying and Haghighi, Aria},
title = {TwHIN: Embedding the Twitter Heterogeneous Information Network for Personalized Recommendation},
year = {2022},
url = {https://doi.org/10.1145/3534678.3539080},
booktitle = {Proceedings of the 28th ACM SIGKDD Conference on Knowledge Discovery and Data Mining}
}

@inproceedings{xiaoKDD20,
author = {Xiao, Zhiping and Song, Weiping and Xu, Haoyan and Ren, Zhicheng and Sun, Yizhou},
title = {TIMME: Twitter Ideology-detection via Multi-task Multi-relational Embedding},
year = {2020},
url = {https://doi.org/10.1145/3394486.3403275},
booktitle = {Proceedings of the 26th ACM SIGKDD International Conference on Knowledge Discovery \& Data Mining}
}

@inproceedings{salemiACL24,
    title = "{L}a{MP}: When Large Language Models Meet Personalization",
    author = "Salemi, Alireza  and
      Mysore, Sheshera  and
      Bendersky, Michael  and
      Zamani, Hamed",
    booktitle = "Proceedings of the 62nd Annual Meeting of the Association for Computational Linguistics (Volume 1: Long Papers)",
    year = "2024",
    url = "https://aclanthology.org/2024.acl-long.399/"
}

@inproceedings{islam2021analysis,
  title={Analysis of {T}witter {U}sers' {L}ifestyle {C}hoices using {J}oint {E}mbedding {M}odel},
  author={Islam, Tunazzina and Goldwasser, Dan},
  booktitle={Proceedings of the International AAAI Conference on Web and Social Media},
  volume={15},
  pages={242--253},
  year={2021}
}

@misc{drukerman24,
      title={The X Types--Mapping the Semantics of the Twitter Sphere}, 
      author={Ogen Schlachet Drukerman and Einat Minkov},
      year={2024},
      eprint={2409.14584},
      archivePrefix={arXiv},
      primaryClass={cs.CL},
      url={https://arxiv.org/abs/2409.14584}, 
}

@inproceedings{wangNAACL24,
    title = "{R}ec{M}ind: Large Language Model Powered Agent For Recommendation",
    author = "Wang, Yancheng  and
      Jiang, Ziyan  and
      Chen, Zheng  and
      Yang, Fan  and
      Zhou, Yingxue  and
      Cho, Eunah  and
      Fan, Xing  and
      Lu, Yanbin  and
      Huang, Xiaojiang  and
      Yang, Yingzhen",
    booktitle = "Findings of the Association for Computational Linguistics: NAACL 2024",
    year = "2024",
    url = "https://aclanthology.org/2024.findings-naacl.271/"
}

@inproceedings{liuACL24,
    title = "Self-Supervised Position Debiasing for Large Language Models",
    author = "Liu, Zhongkun  and
      Chen, Zheng  and
      Zhang, Mengqi  and
      Ren, Zhaochun  and
      Ren, Pengjie  and
      Chen, Zhumin",
    booktitle = "Findings of the Association for Computational Linguistics: ACL 2024",
    year = "2024",
    url = "https://aclanthology.org/2024.findings-acl.170/"
}

@inproceedings{lyuNAACL24,
    title = "{LLM}-Rec: Personalized Recommendation via Prompting Large Language Models",
    author = "Lyu, Hanjia  and
      Jiang, Song  and
      Zeng, Hanqing  and
      Xia, Yinglong  and
      Wang, Qifan  and
      Zhang, Si  and
      Chen, Ren  and
      Leung, Chris  and
      Tang, Jiajie  and
      Luo, Jiebo",
    booktitle = "Findings of the Association for Computational Linguistics: NAACL 2024",
    year = "2024",
    url = "https://aclanthology.org/2024.findings-naacl.39/"
}

@inproceedings{caoNAACL24,
    title = "Aligning Large Language Models with Recommendation Knowledge",
    author = "Cao, Yuwei  and
      Mehta, Nikhil  and
      Yi, Xinyang  and
      Hulikal Keshavan, Raghunandan  and
      Heldt, Lukasz  and
      Hong, Lichan  and
      Chi, Ed  and
      Sathiamoorthy, Maheswaran",
    booktitle = "Findings of the Association for Computational Linguistics: NAACL 2024",
    year = "2024",
    url = "https://aclanthology.org/2024.findings-naacl.67/"
}

@inproceedings{liangColing25,
    title = "Taxonomy-Guided Zero-Shot Recommendations with {LLM}s",
    author = "Liang, Yueqing  and
      Yang, Liangwei  and
      Wang, Chen  and
      Xu, Xiongxiao  and
      Yu, Philip S.  and
      Shu, Kai",
    booktitle = "Proceedings of the 31st International Conference on Computational Linguistics",
    year = "2025",
    url = "https://aclanthology.org/2025.coling-main.102/",
}

@inproceedings{mysore24,
    title = "Pearl: Personalizing Large Language Model Writing Assistants with Generation-Calibrated Retrievers",
    author = "Mysore, Sheshera  and
      Lu, Zhuoran  and
      Wan, Mengting  and
      Yang, Longqi  and
      Sarrafzadeh, Bahareh  and
      Menezes, Steve  and
      Baghaee, Tina  and
      Gonzalez, Emmanuel Barajas  and
      Neville, Jennifer  and
      Safavi, Tara",
    booktitle = "Proceedings of the 1st Workshop on Customizable NLP: Progress and Challenges in Customizing NLP for a Domain, Application, Group, or Individual (CustomNLP4U)",
    month = nov,
    year = "2024",
    url = "https://aclanthology.org/2024.customnlp4u-1.16/"
}

@inproceedings{Richardson2023,
 author = {Chris Richardson and Yao Zhang and Kellen Gillespie and Sudipta Kar and Arshdeep Singh and Zeynab Raeesy and Omar Zia Khan and Abhinav Sethy},
 title = {Integrating summarization and retrieval for enhanced personalization via large language models},
 booktitle = "CIKM 2023 Workshop Personalized Generative AI",
 year = {2023},
 url = {https://www.amazon.science/publications/integrating-summarization-and-retrieval-for-enhanced-personalization-via-large-language-models},
}

@inproceedings{tanEMNLP24,
    title = "Democratizing Large Language Models via Personalized Parameter-Efficient Fine-tuning",
    author = "Tan, Zhaoxuan  and
      Zeng, Qingkai  and
      Tian, Yijun  and
      Liu, Zheyuan  and
      Yin, Bing  and
      Jiang, Meng",
    booktitle = "Proceedings of the 2024 Conference on Empirical Methods in Natural Language Processing",
    year = "2024",
    url = "https://aclanthology.org/2024.emnlp-main.372/"
}

@inproceedings{culotta2015predicting,
  title={Predicting the {D}emographics of {T}witter {U}sers from {W}ebsite {T}raffic {D}ata},
  author={Culotta, Aron and Kumar, Nirmal and Cutler, Jennifer},
  booktitle={Proceedings of the AAAI conference on artificial intelligence},
  volume={29},
  issue={1},
  year={2015}
}

@inproceedings{volkovaACL16,
  title={Inferring perceived demographics from user emotional tone and user-environment emotional contrast},
  author={Volkova, Svitlana and Bachrach, Yoram},
  booktitle={Proceedings of the 54th Annual Meeting of the Association for Computational Linguistics (Volume 1: Long Papers)},
  pages={1567--1578},
  year={2016}
}

@inproceedings{pritskerIUI17,
author = {Wasserman Pritsker, Evgenia and Kuflik, Tsvi and Minkov, Einat},
title = {Assessing the Contribution of Twitter's Textual Information to Graph-based Recommendation},
year = {2017},
url = {https://doi.org/10.1145/3025171.3025218},
booktitle = {Proceedings of the 22nd International Conference on Intelligent User Interfaces}
}

@article{dangur20,
title = {Identification of topical subpopulations on social media},
journal = {Information Sciences},
volume = {528},
pages = {92-112},
year = {2020},
issn = {0020-0255},
url = {https://www.sciencedirect.com/science/article/pii/S0020025520303042},
author = {Ido Dangur and Ron Bekkerman and Einat Minkov}
}

@inproceedings{pendzelEMNLP25,
    title = "Towards Author-informed {NLP}: Mind the Social Bias",
    author = "Pendzel, Inbar and
      Minkov, Einat",
    booktitle = "Proceedings of the 2025 Conference on Empirical Methods in Natural Language Processing",
    year = "2025",
    url = "https://aclanthology.org/2025.emnlp-main.1764/"
}

@article{muellerCSCW21,
  title={Demographic representation and collective storytelling in the me too twitter hashtag activism movement},
  author={Mueller, Aaron and Wood-Doughty, Zach and Amir, Silvio and Dredze, Mark and Nobles, Alicia Lynn},
  journal={Proceedings of the ACM on Human-Computer Interaction},
  volume={5},
  number={CSCW1},
  pages={1--28},
  year={2021},
  publisher={ACM New York, NY, USA}
}

@article{zhangTIST,
author = {Zang, Tianzi and Zhu, Yanmin and Liu, Haobing and Zhang, Ruohan and Yu, Jiadi},
title = {A Survey on Cross-domain Recommendation: Taxonomies, Methods, and Future Directions},
year = {2022},
issue_date = {April 2023},
publisher = {Association for Computing Machinery},
address = {New York, NY, USA},
volume = {41},
number = {2},
url = {https://doi.org/10.1145/3548455},
journal = {ACM Transactions on Information Systems},
month = dec,
articleno = {42}
}

@inproceedings{xuWWW24,
author = {Xu, Wujiang and Wu, Qitian and Wang, Runzhong and Ha, Mingming and Ma, Qiongxu and Chen, Linxun and Han, Bing and Yan, Junchi},
title = {Rethinking Cross-Domain Sequential Recommendation under Open-World Assumptions},
booktitle = {Proceedings of the ACM Web Conference 2024},
year = {2024},
url = {https://doi.org/10.1145/3589334.3645351}
}

@inproceedings{wuICML21,
  title={Towards open-world recommendation: An inductive model-based collaborative filtering approach},
  author={Wu, Qitian and Zhang, Hengrui and Gao, Xiaofeng and Yan, Junchi and Zha, Hongyuan},
  booktitle={International Conference on Machine Learning},
  year={2021}
}

@inproceedings{Mikolov2013,
  author    = {Tom{\'{a}}s Mikolov and Kai Chen and Greg Corrado and Jeffrey Dean},
  title     = {Efficient Estimation of Word Representations in Vector Space},
  booktitle = {1st International Conference on Learning Representations, {ICLR}},
  year      = {2013}
}

@inproceedings{jiSIGIR20,
author = {Ji, Yitong and Sun, Aixin and Zhang, Jie and Li, Chenliang},
title = {A Re-visit of the Popularity Baseline in Recommender Systems},
year = {2020},
url = {https://doi.org/10.1145/3397271.3401233},
booktitle = {Proceedings of the 43rd International ACM SIGIR Conference on Research and Development in Information Retrieval}
}

@inproceedings{sutterEACL24,
    title = "Unsupervised stance detection for social media discussions: A generic baseline",
    author = "Sutter, Maia  and
      Gourru, Antoine  and
      Trabelsi, Amine  and
      Largeron, Christine",
    booktitle = "Proceedings of the 18th Conference of the European Chapter of the Association for Computational Linguistics (Volume 1: Long Papers)",
    month = mar,
    year = "2024",
    publisher = "Association for Computational Linguistics",
    url = "https://aclanthology.org/2024.eacl-long.107/"
}

\end{document}